\documentclass[twocolumn,journal]{IEEEtran}
\usepackage[T1]{fontenc}
\usepackage{bm}
\usepackage{amsthm}
\usepackage{amsmath}
\usepackage{amssymb}
\usepackage{graphicx}
\usepackage[unicode=true,
 bookmarks=true,bookmarksnumbered=true,bookmarksopen=true,bookmarksopenlevel=1,
 breaklinks=false,pdfborder={0 0 0},backref=false,colorlinks=false]
 {hyperref}
\hypersetup{pdftitle={Your Title},
 pdfauthor={Your Name},
 pdfpagelayout=OneColumn, pdfnewwindow=true, pdfstartview=XYZ, plainpages=false}
\usepackage{breakurl}

\makeatletter

\providecommand{\tabularnewline}{\\}

\theoremstyle{plain}
\newtheorem{thm}{\theoremname}
\theoremstyle{definition}
\newtheorem{defn}{\definitionname}
\theoremstyle{remark}
\newtheorem{rem}{\remarkname}
\theoremstyle{plain}
\newtheorem{lem}{\lemmaname}
\theoremstyle{plain}
\newtheorem{prop}{\propositionname}

\usepackage[vlined,boxed,ruled,linesnumbered,resetcount]{algorithm2e}
\usepackage{algpseudocode}

\usepackage[noadjust,sort,compress]{cite}
\usepackage{epstopdf}
\makeatother

\providecommand{\definitionname}{Definition}
\providecommand{\lemmaname}{Lemma}
\providecommand{\propositionname}{Proposition}
\providecommand{\remarkname}{Remark}
\providecommand{\theoremname}{Theorem}

\begin{document}

\title{Hybrid Vector Perturbation Precoding: The Blessing of Approximate
Message Passing}

\author{Shanxiang Lyu and~Cong Ling,~\IEEEmembership{Member,~IEEE}\thanks{S. Lyu and C. Ling are with the Department of Electrical and Electronic
Engineering, Imperial College London, London SW7 2AZ, United Kingdom
(e-mail: s.lyu14@imperial.ac.uk, cling@ieee.org). }}
\maketitle
\begin{abstract}
Vector perturbation (VP) precoding is a promising technique for multiuser
communication systems operating in the downlink. In this work, we
introduce a hybrid framework to improve the performance of lattice
reduction (LR) aided precoding in VP. First, we perform a simple precoding
using zero forcing (ZF) or successive interference cancellation (SIC)
based on a reduced lattice basis. Since the signal space after LR-ZF
or LR-SIC precoding can be shown to be bounded to a small range, then
along with sufficient orthogonality of the lattice basis guaranteed
by LR, they collectively pave the way for the subsequent application
of an approximate message passing (AMP) algorithm, which further boosts
the performance of any suboptimal precoder. Our work shows that the
AMP algorithm can be beneficial for a lattice decoding problem whose
data symbols lie in integers $\mathbb{Z}$ and entries of the lattice
basis may bot be i.i.d. Gaussian. Numerical results confirm the low-complexity
AMP algorithm can improve the symbol error rate (SER) performance
of LR aided precoding significantly. Lastly, the hybrid scheme is
also proved effective when solving the data detection problem of massive
MIMO systems without using LR.\end{abstract}


\section{Introduction}

\noindent \IEEEPARstart{T}{he } broadband mobile internet of the
next generation is expected to deliver high volume data to a large
number of users simultaneously. To meet this demand in the multiuser
broadcast network, it is desirable to precode the transmit symbols
according to the channel state information (CSI) with improved time-efficiency
while retaining the reliability. It is known that precoding by using
plain channel inversion performs poorly at all singal-to-noise ratios
(SNRs), and further regularization cannot improve the performance
substantially. To enhance the throughput, a precoding scheme called
vector perturbation (VP) was introduced in \cite{Peel2005,Hochwald2005}.
The scheme is based on Tomlinson-Harashima precoding which perturbs
the transmitted data by modulo-lattice operations, and it can achieve
near-sum-capacity of the system without using dirty-paper techniques
\cite{Peel2005,Hochwald2005}. The optimization target of VP requires
to solve the closest vector problem (CVP) in a lattice, which has
been proved NP-complete by a reduction from the decision version of
CVP \cite{Micciancio2002}. Due to the NP-complete nature of the problem,
finding its exact solution using sphere decoding \cite{Agrell2002}
(referred to as sphere precoding in \cite{Peel2005,Hochwald2005})
incurs a prohibitive computational complexity that grows exponentially
with the dimension of the problem. Therefore, reduced-complexity alternatives
providing near-optimal performance must suffice.

Several reduced-complexity precoding algorithms have been proposed
in the literature \cite{Masouros2013,Han2010,Park2011a,Liu2012a,Masouros2014,Karpuk2016,Ma2016a}.
These algorithms are split into two categories based on whether lattice
reduction has been used as pre-processing. In the first category \cite{Masouros2013,Han2010,Park2011a,Ma2016a,Masouros2014},
decoding of CVP is solved on the original input basis, and the advantages
of low complexity is due to the constraints imposed on the signal
space (c.f. \cite{Han2010,Park2011a}) or the lattice basis (c.f.
\cite{Masouros2013,Masouros2014}). There is however no theoretical
performance guarantee for these simplified methods, so we have to
resort to approaches in the second category \cite{Liu2012a,Windpassinger2004,Taherzadeh2007a}.
These approaches are referred to as lattice reduction (LR) aided precoding
(decoding), which consists of lattice reduction as pre-processing
and approximated decoding using zero-forcing (ZF), successive interference
cancellation (SIC) or other variants. Thanks to the good properties
of a reduced basis, approximated decoding based on it has been shown
to achieve full diversity order \cite{Taherzadeh2007a,Liu2012a}.
Compared to algorithms in the first category, the pre-processing complexity
of reducing a lattice basis varies from being polynomial to exponential
(cf. \cite{Lenstra1982,Lyu2017}). This cost is however not an issue
\cite{Taherzadeh2007a} in slow-fading channels where the lattice
basis is fixed during a large number of time slots, because the lattice
basis is only reduced once to serve all the CVP instances.

Focusing on the framework with LR, the aim of this paper is to design
a low-complexity message passing algorithm after the phase of approximated
decoding. The fundamental principle of message passing algorithms
is to decompose high-dimensional problems into sets of smaller low-dimensional
problems. This decomposition is often interpreted in a bipartite graph,
where the problem variables and factors are represented by graph vertices
and dependencies between them represented by edges. Exact message
passing methods such as belief propagation (BP) \cite{Weiss2001,Richardson2008}
exploit this graphical structure to perform optimization in an iterative
manner. By simplifying BP, a new class of low-complexity iterative
algorithms referred to as AMP was proposed in \cite{Donoho2009,Donoho2009a},
and rigorous justification on their performance can be found in \cite{Bayati2011,Bayati2015}. 

Inspired by the applications \cite{Jeon2015,Jeon2016,Liu2016a,Chen2017}
of approximate message passing (AMP) \cite{Donoho2009,Donoho2009a}
in data detection of massive multiple-input multiple-output (MIMO)
systems, we investigate a general issue of how to use AMP to solve
CVP. By saying general, we emphasize that the data symbols to be estimated
in message passing reside in integers $\mathbb{Z}$ which are infinite.
As Bayati and Montanari \cite{Bayati2011} had mentioned their state
evolution analysis of AMP can extend beyond compressed sensing (to
linear estimation and multi-user detection), we may wonder why AMP
cannot be adopted for CVP in a straightforward manner. Actually, even
assuming the data symbols are only taken from a finite discrete constellation
already complicates the problem of using AMP. For instance, \cite{Jeon2015}
showed that the channel matrix has to become extremely tall as the
size the the constellation grows, and \cite{Jeon2016} argued that
the calculation of the posterior mean function of AMP becomes numerically
unstable for small values of noise variance. We also noticed that
the posterior mean function (denoted as threshold function in \cite{Bayati2011})
is not Lipschitz continuous for small values of noise variance, so
the theoretical justification of AMP does not hold in this scenario
(c.f. \cite[Section 2.3]{Bayati2011}). Although we may bypass these
issues by using Gaussian distributions as mismatched data distributions,
as used in \cite{lyu15,Jeon2016,Liu2016a,Chen2017}, it is easily
recognized that their performance is no better than that of Linear
Minimum Mean Square Error (LMMSE) estimation. To embrace the low-complexity
advantage of AMP and to address the aforementioned issues, it motivates
us to design a new decoding architecture for CVP. The key results
and contributions of our work are summarized as follows.

1) We propose a hybrid precoding scheme, which uses AMP in conjunction
with a sub-optimal estimator after lattice reduction. Considering
the theoretical properties and practical performance, we choose the
sub-optimal estimator as ZF or SIC, and set the lattice reduction
methods as boosted versions of Lenstra\textendash Lenstra\textendash Lovász
(b-LLL) or Korkine-Zolotarev (b-KZ) \cite{Lenstra1982,Korkinge1877,Lyu2017}.
After that, we analyze the energy efficiency of precoding with LR-ZF/LR-SIC.
On the basis of the proved upper bounds on the energy efficiency,
we can deduce upper bounds for the range of data symbols to be estimated
by AMP. Since these bounds are derived from a worst case analysis,
we also study their empirical distributions. 

2) As a reduced lattice basis may not have uniform power in all the
columns, we use the approximation techniques in \cite{Montanari2010,Maleki2011}
to derive the corresponding AMP algorithm based on simplifying BP.
The underlying state evolution equation of it is derived. Subsequently
we propose to use ternary distributions and Gaussian distributions
for the threshold functions in AMP, whose posterior mean and variance
functions have closed-form expressions. The impacts of a reduced basis
and parameters in the chosen prior distributions are studied based
on the state evolution equation. Simulation results reveal that concatenating
AMP to LR-ZF/LR-SIC can provide significant performance improvements.

3) After solving the underlying CVP in VP, the corresponding CVP in
massive MIMO can also be solved in an easier manner. Specifically,
the lattice bases (channel matrices) in the uplink data detection
problem of massive MIMO systems are naturally short and orthogonal,
so it suggests we can apply the hybrid scheme to this scenario without
using lattice reduction. Simulation results confirm the effectiveness
of this extension.\textit{ }

The rest of this paper is organized as follows. We review some basic
concepts about lattices and VP in Section \ref{sec:Preliminaries}.
The hybrid scheme is explained in Section \ref{sec:The-hybrid-scheme},
which includes demonstrations about why we have reached another problem
with a finite constellation size. Section \ref{sec:Our-AMP-L-algorithm}
presents our AMP algorithm. Simulation results for VP are given in
Section \ref{secamp:Simulations}. The extension to massive MIMO is
presented in Section \ref{sec:Extension-to-Data}, and the last section
concludes this paper. 

Notations: Matrices and column vectors are denoted by uppercase and
lowercase boldface letters. We use $\mathbb{R}$ and $\mathbb{Z}$
to represent the field of real numbers and the ring of rational integers,
respectively. $\mathrm{GL}_{n}\left(\mathbb{Z}\right)$ refers to
a general linear group with entries in $\mathbb{Z}$. $\lfloor\cdot\rceil$,
$|\cdot|$ and $\left\Vert \cdot\right\Vert $ respectively refer
to (element-wise) rounding, taking the absolute value, and taking
the Euclidean norm. $H_{i,j}$ denotes the $(i,j)$th entry of matrix
$\mathbf{H}$. $\mathbf{H}^{\top}$ and $\mathbf{H}^{\dagger}=\left(\mathbf{H}^{\top}\mathbf{H}\right)^{-1}\mathbf{H}^{\top}$
denote the transpose and the Moore-Penrose pseudo-inverse of $\mathbf{H}$,
respectively. $\mathrm{span}(\mathbf{S})$ denotes the vector space
spanned by $\mathbf{S}$. $\pi_{\mathbf{S}}(\mathbf{x})$ and $\pi_{\mathbf{S}}^{\bot}(\mathbf{x})$
denote the projection of $\mathbf{x}$ onto $\mathrm{span}(\mathbf{S})$
and the orthogonal complement of $\mathrm{span}(\mathbf{S})$, respectively.
$\propto$ stands for equality up to a normalization constant. $[n]$
denotes $\left\{ 1,\ldots\thinspace,n\right\} $, $\langle\mathbf{x}\rangle=\sum_{j=1}^{n}x_{j}/n$.
$\mathrm{N}(\mu,\Sigma)$ represents a multivariate normal distribution
with mean $\mu$ and covariance matrix $\Sigma$. We use the standard
asymptotic notation $p(x)=O(q(x))$ when $\lim\sup_{x\rightarrow\infty}|p(x)/q(x)|<\infty$.

\section{\label{sec:Preliminaries}Preliminaries}

\subsection{Lattices}

An $n$-dimensional lattice is a discrete additive subgroup in $\mathbb{R}^{n}$.
A $\mathbb{Z}$-lattice with basis $\mathbf{H}=[\mathbf{h}_{1},\ldots\thinspace,\mathbf{h}_{n}]\in\mathbb{R}^{m\times n}$
can be represented by 
\[
\mathcal{L}(\mathbf{H})=\left\{ \mathbf{v}\mid\mathbf{v}=\sum_{i\in[n]}c_{i}\mathbf{h}_{i},\thinspace c_{i}\in\mathbb{Z}\right\} .
\]

It is necessary to know whether the basis vectors $\mathbf{h}_{i}$'s
are short and nearly orthogonal. This property can be measured by
the orthogonality defect (OD): 
\begin{equation}
\xi(\mathbf{H})=\frac{\prod_{i=1}^{n}\left\Vert \mathbf{h}_{i}\right\Vert }{\sqrt{\det(\mathbf{H}^{\top}\mathbf{H})}}.\label{eq:OD}
\end{equation}
We have $\xi(\mathbf{H})\geq1$ due to Hadamard's inequality. Given
$\mathbf{H}$, the denominator of (\ref{eq:OD}) is fixed, while the
$\left\Vert \mathbf{h}_{i}\right\Vert $ in the numerator can be reduced
to get close to the $i$th successive minimum of $\mathcal{L}(\mathbf{H})$,
which is defined by the smallest real number $r$ such that $\mathcal{L}(\mathbf{H})$
contains $i$ linearly independent vectors of length at most $r$:
\[
\lambda_{i}(\mathbf{H})=\inf\left\{ r\mid\dim(\mathrm{span}((\mathcal{L}\cap\mathcal{B}(\mathbf{0},r)))\geq i\right\} ,
\]
in which $\mathcal{B}(\mathbf{0},r)$ denotes a ball centered at the
origin with radius $r$. 

The goal of lattice reduction is to find, for a given lattice, a basis
matrix with favorable properties. There are many well developed reduction
algorithms. Here we review the polynomial time LLL \cite{Lenstra1982}
reduction and the exponential time KZ \cite{Korkinge1877} reduction,
followed by their boosted variants. 
\begin{defn}[\cite{Lenstra1982}]
A basis $\mathbf{H}$ is called LLL reduced if it satisfies the size
reduction conditions of $|R_{i,j}/R_{i,i}|\leq\frac{1}{2}$ for $1\leq i\leq n$,
$j>i$, and Lovász's conditions of $\delta R_{i,i}^{2}\leq R_{i,i+1}^{2}+R_{i+1,i+1}^{2}$
for $1\leq i\leq n-1$.
\end{defn}
In the definition, $R_{i,j}$'s refer to elements of the $\mathbf{R}$ matrix
of the QR decomposition on $\mathbf{H}$, and $\delta\in(1/4,1)$
is called Lovász's constant. Define $\beta=1/\sqrt{\delta-1/4}\in(2/\sqrt{3},\infty)$,
for an LLL reduced basis $\mathbf{H}$ we have \cite{Lenstra1982}
\begin{equation}
\xi(\mathbf{H})\leq\beta^{n(n-1)/2}.\label{eq:lll bound}
\end{equation}

\begin{defn}[\cite{Lagarias1990}]
A basis $\mathbf{H}$ is called KZ reduced if it satisfies the size
reduction conditions, and the projection conditions of $\pi_{\left[\mathbf{h}_{1},\ldots\thinspace,\mathbf{h}_{i-1}\right]}^{\perp}(\mathbf{h}_{i})$
being the shortest vector of the projected lattice $\pi_{\left[\mathbf{h}_{1},\ldots\thinspace,\mathbf{h}_{i-1}\right]}^{\perp}([\mathbf{h}_{i},\ldots\thinspace,\mathbf{h}_{n}])$
for $1\leq i\leq n$. 
\end{defn}
If $\mathbf{H}$ is KZ reduced, we have \cite{Lagarias1990} 
\begin{equation}
\xi(\mathbf{H})\leq\left(\prod_{i=1}^{n}\frac{\sqrt{i+3}}{2}\right)\left(\frac{2n}{3}\right)^{n/2}.\label{eq:HKZ bound}
\end{equation}
In this paper we will adopt the boosted version of LLL/KZ so as to
get shorter and more orthogonal basis vectors \cite{Lyu2017} .
\begin{defn}[\cite{Lyu2017}]
A basis $\mathbf{H}$ is called boosted LLL (b-LLL) reduced if it
satisfies diagonal reduction conditions of $\delta R_{i,i}^{2}\leq(R_{i,i+1}-\lfloor R_{i,i+1}/R_{i,i}\rceil R_{i,i})^{2}+R_{i+1,i+1}^{2}$
for $1\leq i\leq n-1$, and all $\mathbf{h}_{i}$ for $2\leq i\leq n$
are reduced by an approximate CVP oracle with list size $p$ along
with a rejection operation.

Although the definition of b-LLL ensures that it performs no worse
than LLL in reducing the lengths of basis vectors, only the same bound
on OD has been proved: $\xi(\mathbf{H})\leq\beta^{n(n-1)/2}$ \cite{Lyu2017}.
\end{defn}

\begin{defn}[\cite{Lyu2017}]
A basis $\mathbf{H}$ is called boosted KZ (b-KZ) reduced if it satisfies
the projection conditions as KZ, and the length reduction conditions
of $\left\Vert \mathbf{h}_{i}\right\Vert \leq\left\Vert \mathbf{h}_{i}-\mathcal{Q}_{\mathcal{L}(\left[\mathbf{h}_{1},\ldots\thinspace,\mathbf{h}_{i-1}\right])}(\pi_{\left[\mathbf{h}_{1},\ldots\thinspace,\mathbf{h}_{i-1}\right]}(\mathbf{h}_{i}))\right\Vert $
for $2\leq i\leq n$, where $\mathcal{Q}_{\mathcal{L}(\left[\mathbf{h}_{1},\ldots\thinspace,\mathbf{h}_{i-1}\right])}\left(\cdot\right)$
is the nearest neighbor quantizer w.r.t. $\mathcal{L}(\left[\mathbf{h}_{1},\ldots\thinspace,\mathbf{h}_{i-1}\right])$.
\end{defn}
If $\mathbf{H}$ is b-KZ reduced, we have 
\begin{equation}
\xi(\mathbf{H})\leq\frac{\sqrt{n}}{2}\left(\prod_{i=1}^{n-1}\frac{\sqrt{i+3}}{2}\right)\left(\frac{2n}{3}\right)^{n/2}.\label{eq:HKZ bound-1}
\end{equation}

\subsection{Vector Perturbation and CVP }

Vector perturbation is a non-linear precoding technique that aims
to minimize the transmitted power that is associated with the transmission
of a certain data vector \cite{Peel2005,Hochwald2005}. Assume the
base station is equipped with $m$ transmit antennas to broadcast
messages to $n$ individual users, and each user has only one antenna.
The observed signals at users $1$ to $n$ can be collectively expressed
as as a vector: 
\begin{equation}
\bar{\mathbf{t}}=\mathbf{B}\mathbf{t}+\bar{\mathbf{w}}\label{eq:1}
\end{equation}
 where $\mathbf{B}\in\mathbb{R}^{n\times m}$ denotes a channel matrix
whose entries admit $\mathrm{N}(0,1)$, $\mathbf{t}\in\mathbb{R}^{m}$
is a transmitted signal, and $\bar{\mathbf{w}}\sim\mathrm{N}(\mathbf{0},\sigma_{w}^{2}\mathbf{I}_{n})$
denotes additive Gaussian noise. 

With perfect channel knowledge at the base station, the transmitted
signal $\mathbf{t}$ is designed to be a truncation of the channel
inversion precoding $\mathbf{B}^{\dagger}\mathbf{s}$: 
\begin{equation}
\mathbf{t}=\mathbf{B}^{\dagger}(\mathbf{s}-M\mathbf{x}),\label{eq:2}
\end{equation}
where $\mathbf{x}\in\mathbb{Z}^{n}$ is an integer vector to be optimized,
$\mathbf{s}\in\mathcal{M}^{n}$ is the symbol vector. We set the constellation
as $\mathcal{M}=\{0,\ldots,M-1\}$ where $M>1$ is a positive integer.
All quadrature amplitude modulation (QAM) constellations can be transformed
to this format after adjusting (\ref{eq:2}).

Assume the transmitted signal has unit power, and let $E_{\mathbf{t}}\triangleq\left\Vert \mathbf{t}\right\Vert $
be a normalization factor. Then the signal vector at users is represented
by
\begin{equation}
\bar{\mathbf{t}}=(\mathbf{s}-M\mathbf{x})/E_{\mathbf{t}}+\bar{\mathbf{w}}.\label{eq:3}
\end{equation}
Let $\bar{\mathbf{t}}'=E_{\mathbf{t}}\bar{\mathbf{t}}$, $\bar{\mathbf{w}}'=E_{\mathbf{t}}\bar{\mathbf{w}}$,
since $M\mathbf{x}\thinspace\mod\thinspace M=\mathbf{0}$, the above
equation can be transformed to 
\begin{equation}
\lfloor\bar{\mathbf{t}}'\rceil\thinspace\mod\thinspace M=\lfloor\mathbf{s}+\bar{\mathbf{w}}'\rceil\thinspace\mod\thinspace M.\label{eq:recover1}
\end{equation}
From (\ref{eq:recover1}), we can see that if $|\bar{w}_{i}'|<\frac{1}{2}$
$\forall\thinspace i$, where $\bar{\mathbf{w}}'\in\mathrm{N}(\bm{0},\sigma_{w}^{2}E_{\mathbf{t}}\mathbf{I}_{n})$,
then $\mathbf{s}$ can be faithfully recovered.

To decrease the decoding error probability which is dominated by $E_{\mathbf{t}}$,
the transmitter has to address the following optimization problem:
\begin{equation}
\arg\min_{\mathbf{x}\in\mathbb{Z}^{n}}\left\Vert \mathbf{B}^{\dagger}(\mathbf{s}-M\mathbf{x})\right\Vert ^{2}.\label{eq:beforesvp}
\end{equation}
Define $\mathbf{y}=\mathbf{B}^{\dagger}\mathbf{s}\in\mathbb{R}^{m}$,
$\mathbf{H}=M\mathbf{B}^{\dagger}\in\mathbb{R}^{m\times n}$, then
(\ref{eq:beforesvp}) represents a CVP instance of lattice decoding:
\begin{equation}
\mathbf{x}^{\mathrm{cvp}}=\arg\min_{\mathbf{x}\in\mathbb{Z}^{n}}\left\Vert \mathbf{y}-\mathbf{H}\mathbf{x}\right\Vert ^{2}.\label{eq:CVP}
\end{equation}

This CVP is different from the CVP in MIMO detection because the distance
distribution from $\mathbf{y}$ to lattice $\mathcal{L}(\mathbf{H})$
is not known, the lattice basis is the inverse of the channel matrix
that has highly correlated entries, and the data symbols are optimized
over $\mathbb{Z}^{n}$ rather than over a small finite constellation.

\section{\label{sec:The-hybrid-scheme}The hybrid precoding scheme}

Our hybrid precoding scheme to solve the CVP in (\ref{eq:CVP}) consists
of two phases: 

\begin{figure}[th]
\center

\includegraphics[clip,width=3.4in]{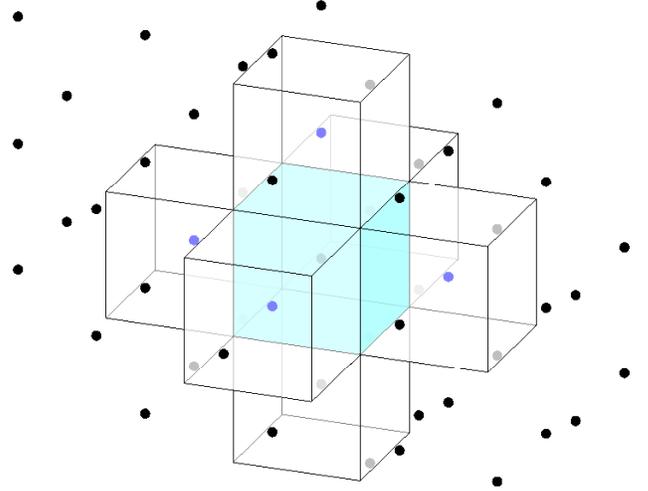}

\protect\protect\caption{Exploring the vicinity of a good candidate $\mathbf{x}^{\mathrm{zf}}\in\mathbb{R}^{3}$,
whose decision parallelepiped is the cyan cube. After updating the
target vector $\mathbf{y\leftarrow y}-\mathbf{H}\mathbf{x}^{\mathrm{zf}}$,
to optimize $\min_{\mathbf{x}\in\left\{ -1,0,1\right\} ^{3}}\left\Vert \mathbf{y}-\mathbf{H}\mathbf{x}\right\Vert $
enables locating all the blue lattice points inside the white cubes.}
\label{fig lattice space} 
\end{figure}

\textbf{Phase 1} (\textbf{approximated decoding}): Apply lattice reduction
to the input basis to get $\mathbf{H}\leftarrow\mathbf{H}\mathbf{U}$,
where $\mathbf{U}\in\mathrm{GL}_{n}(\mathbb{Z})$ is induced by the
reduction operation. Based on the reduced $\mathbf{H}$, use ZF or
SIC to get a sub-optimal candidate: $\hat{\mathbf{x}}=\mathbf{x}^{\mathrm{zf}}$
or $\hat{\mathbf{x}}=\mathbf{x}^{\mathrm{sic}}$. 

\textbf{Phase 2} (\textbf{AMP decoding}): Let $\mathbf{y}\leftarrow\mathbf{y}-\mathbf{H}\hat{\mathbf{x}}$
and define a finite constraint $\mathcal{B}=\left\{ -B,-B+1,\ldots,B-1,B\right\} $.
After that, use an AMP algorithm to solve: 
\begin{equation}
\mathbf{x}^{\mathrm{amp}}=\arg\min_{\mathbf{x}\in\mathcal{B}^{n}}\left\Vert \mathbf{y}-\mathbf{H}\mathbf{x}\right\Vert ^{2}.\label{eq:AMP MODEL}
\end{equation}
Lastly return $\hat{\mathbf{x}}+\mathbf{x}^{\mathrm{amp}}$.

The underlying rationale is demonstrated in Fig. \ref{fig lattice space}.
Regarding the algorithmic routines in Phase 1, $\mathbf{x}^{\mathrm{zf}}=\lfloor\mathbf{H}^{\dagger}\mathbf{y}\rceil$,
and we refer to \cite{Nguyen2010,Lyu2017} for those of lattice reduction,
to \cite{Ling2011} for that of $\mathbf{x}^{\mathrm{sic}}$.

To ensure the hybrid decoding scheme is correct and efficient, the
following two issues are addressed in the paper.
\begin{itemize}
\item Regarding the transformation from (\ref{eq:CVP}) to (\ref{eq:AMP MODEL}),
one has to specify a minimum range for constraint $\mathcal{B}^{n}$
such that 
\[
\arg\min_{\mathbf{x}\in\mathcal{B}^{n}}\left\Vert \mathbf{y}-\mathbf{H}\hat{\mathbf{x}}-\mathbf{H}\mathbf{x}\right\Vert ^{2}=\arg\min_{\mathbf{x}\in\mathbb{Z}^{n}}\left\Vert \mathbf{y}-\mathbf{H}\mathbf{x}\right\Vert ^{2}.
\]
Generally speaking, problem (\ref{eq:AMP MODEL}) becomes easier if
$B$ is smaller. In Section \ref{sub:The-boundth}, we will examine
the theoretical and empirical bounds of constraint $\mathcal{B}^{n}$.
\item The AMP algorithms in \cite{Donoho2009,Donoho2009a} were assuming
at least the entries of $\mathbf{H}$ being sub-Gaussian with variance
$O(1/n)$. Can we derive an AMP algorithm that is suitable for problem
(\ref{eq:AMP MODEL}), and possibly the routines are simple and have
closed-form expressions? We will first address relevant prerequisites
in Section \ref{sub:Prerequisites-for-AMP}. Considering all the constraints,
Section \ref{sec:Our-AMP-L-algorithm} will present an AMP algorithm
based on simplifying BP.
\end{itemize}

\subsection{\label{sub:The-boundth}The bounds of constraint $\mathcal{B}^{n}$}

In the application to precoding, we show in this section that the
estimation range $\mathcal{B}^{n}$ is bounded after LR-ZF/LR-SIC.
Now we introduce a parameter called energy efficiency to describe
how far a suboptimal perturbation is from the optimal one. 
\begin{defn}
The energy efficiency of an algorithm providing $\hat{\mathbf{x}}$
is the smallest $\eta_{n}$ in the constraint 
\begin{equation}
\left\Vert \mathbf{y}-\mathbf{H}\hat{\mathbf{x}}\right\Vert \leq\eta_{n}\left\Vert \mathbf{y}-\mathbf{H}\mathbf{x}^{\mathrm{cvp}}\right\Vert ,\label{eq:etaCVP}
\end{equation}
where $\mathbf{x}^{\mathrm{cvp}}=\arg\min_{\mathbf{x}\in\mathbb{Z}^{n}}\left\Vert \mathbf{y}-\mathbf{Hx}\right\Vert $,
and we say this algorithm solves $\eta_{n}$-CVP \footnote{In \cite{Liu2012a}, $\eta_{n}$ is referred to as proximity factor
in the CVP context. To avoid confusion with the proximity factor in
\cite{Ling2011}, we simply call it ``energy efficiency''. }.
\end{defn}

We first analyze the energy efficiency $\eta_{n}$ of b-LLL/b-KZ aided
ZF/SIC, and then address the bound for $\mathcal{B}^{n}$ based on
$\eta_{n}$ . The reasons for choosing b-LLL/b-KZ as the reduction
method are: i) b-LLL provides better practical performance than that
of LLL \cite{Lyu2017}, and ii) b-KZ characterizes the theoretical
limit of strong (with exponential complexity) LR methods. 
\begin{thm}
\label{thm:thmlll}For the SIC estimator, if the lattice basis is
reduced by b-LLL, then 
\begin{equation}
\eta_{n}=\beta^{n}/\sqrt{\beta^{2}-1},\label{eq: lllsiz}
\end{equation}
where $\beta\in(2/\sqrt{3},\infty)$; and if the basis is reduced
by b-KZ, then

\begin{equation}
\eta_{n}=1+\frac{8n}{9}\left(n-1\right)^{1+\ln(n-1)/2}.\label{eq:kzsic}
\end{equation}
\end{thm}
\begin{IEEEproof}
The proof relies on upper bounding the diagonal entries of $\mathbf{R}$
(the R matrix in the QR factorization of \textbf{$\mathbf{H}$}).
Since boosted LLL/KZ has the same diagonal entries as those of LLL/KZ,
we can use results about energy efficiency from classic LLL/KZ if
they exist. Hence Eq. (\ref{eq: lllsiz}) is adapted from LLL in \cite{Liu2012a}.
As no such result is known for KZ, we prove a sharp bound for both
KZ and b-KZ in Appendix \ref{sec: short quan}, where the skill involved
is essentially due to \cite{Babai1986}. 
\end{IEEEproof}

\begin{thm}
\label{thm:thmZFlr}For the ZF estimator, if the lattice basis is
reduced by b-LLL, then

\begin{equation}
\eta_{n}=2n\prod_{j=1}^{n}\beta^{j-1}+1;\label{eq:lllzf}
\end{equation}
and if the basis is reduced by b-KZ, then

\begin{equation}
\eta_{n}=2n\prod_{j=1}^{n}j^{2+\ln(j)/2}+1.\label{eq:kzzf}
\end{equation}
\end{thm}
\begin{IEEEproof}
See Appendix \ref{sec:Proof-of-zflr}.
\end{IEEEproof}

Notice that the maximal range of $\mathcal{B}$ is $\max_{i\in n}|\hat{x}_{i}-x_{i}^{\mathrm{cvp}}|$.
Here, we upper bound it by a function about the energy efficiency
$\eta_{n}$. Denote $\varrho(\mathbf{H})$ as the covering radius
of lattice $\mathcal{L}(\mathbf{H})$, it follows from the triangle
inequality and $\left\Vert \mathbf{y}-\mathbf{H}\mathbf{x}^{\mathrm{cvp}}\right\Vert \leq\varrho(\mathbf{H})$
that 
\begin{align*}
\left\Vert \mathbf{H}\left(\hat{\mathbf{x}}-\mathbf{x}^{\mathrm{cvp}}\right)\right\Vert  & \leq\left\Vert \mathbf{y}-\mathbf{H}\hat{\mathbf{x}}\right\Vert +\left\Vert \mathbf{y}-\mathbf{H}\mathbf{x}^{\mathrm{cvp}}\right\Vert \\
 & \leq\left(\eta_{n}+1\right)\varrho(\mathbf{H}).
\end{align*}
 With unitary transform, we have $\left\Vert \mathbf{H}\left(\hat{\mathbf{x}}-\mathbf{x}^{\mathrm{cvp}}\right)\right\Vert =\left\Vert \mathbf{R}\left(\hat{\mathbf{x}}-\mathbf{x}^{\mathrm{cvp}}\right)\right\Vert $.
To get the upper bound for each $|\hat{x}_{i}-x_{i}^{\mathrm{cvp}}|$,
we can expand the quadratic form in the l.h.s. of 
\[
\left\Vert \mathbf{R}\left(\hat{\mathbf{x}}-\mathbf{x}^{\mathrm{cvp}}\right)\right\Vert ^{2}\leq\left(\eta_{n}+1\right)^{2}\varrho^{2}(\mathbf{H})
\]
to get 
\begin{align*}
R_{n,n}^{2}\left(\hat{x}_{n}-x_{n}^{\mathrm{cvp}}\right)^{2}+\cdots+\left(R_{1,1}\left(\hat{x}_{n}-x_{n}^{\mathrm{cvp}}\right)+R_{1,n}\left(\hat{x}_{n}-x_{n}^{\mathrm{cvp}}\right)\right)^{2}\\
\leq\left(\eta_{n}+1\right)^{2}\varrho^{2}(\mathbf{H}).
\end{align*}
For a reduced basis, we know that all the column vectors are short
and the diagonal entries are not very small w.r.t. the successive
minima: $\left\Vert \mathbf{h}_{i}\right\Vert \leq\omega_{i}\lambda_{i}(\mathbf{H})$,
$|R_{i,i}|\geq\lambda_{1}(\mathbf{H})/\varpi_{i}$, where the values
of $\omega_{i}$ and $\varpi_{i}$ can be found in \cite{Lyu2017}.
Now regarding the bound of $|\hat{x}_{n}-x_{n}^{\mathrm{cvp}}|$,
it follows from $R_{n,n}^{2}\left(\hat{x}_{n}-x_{n}^{\mathrm{cvp}}\right)^{2}\leq\left(\eta_{n}+1\right)^{2}\varrho^{2}(\mathbf{H})$
that 
\begin{align*}
|\hat{x}_{n}-x_{n}^{\mathrm{cvp}}| & \leq\left(\eta_{n}+1\right)\varrho(\mathbf{H})/|R_{n,n}|\\
 & \leq\left(\eta_{n}+1\right)\varrho(\mathbf{H})\varpi_{n}/\lambda_{1}(\mathbf{H}).
\end{align*}
Similarly for $|\hat{x}_{n-1}-x_{n-1}^{\mathrm{cvp}}|$, one has
\begin{align*}
|\hat{x}_{n-1}-x_{n-1}^{\mathrm{cvp}}| & \leq\left\Vert \mathbf{R}_{:,1:n-1}\left(\hat{\mathbf{x}}_{1:n-1}-\mathbf{x}_{1:n-1}^{\mathrm{cvp}}\right)\right\Vert /|R_{n-1,n-1}|\\
 & \leq\left(\left(\eta_{n}+1\right)\varrho(\mathbf{H})+\omega_{n}\lambda_{n}(\mathbf{H})|\hat{x}_{n}-x_{n}^{\mathrm{cvp}}|\right)\varpi_{n-1}/\lambda_{1}(\mathbf{H})\\
 & \leq\left(\eta_{n}+1\right)\varpi_{n-1}\varrho(\mathbf{H})/\lambda_{1}(\mathbf{H})\\
 & \thinspace\thinspace\thinspace+\omega_{n}\left(\eta_{n}+1\right)\varpi_{n}\varpi_{n-1}\lambda_{n}(\mathbf{H})\varrho(\mathbf{H})/\lambda_{1}^{2}(\mathbf{H}).
\end{align*}
By induction, we can obtain the upper bounds of $|\hat{x}_{n-2}-x_{n-2}^{\mathrm{cvp}}|,\ldots\thinspace,|\hat{x}_{1}-x_{1}^{\mathrm{cvp}}|$.
The concrete values of these bounds can be evaluated by using the
values of $\eta_{i}$, $\omega_{i}$ and $\varpi_{i}$ based on the
chosen LR aided ZF/SIC algorithms. In summary, the maximal error distance
$\max_{i\in n}|\hat{x}_{i}-x_{i}^{\mathrm{cvp}}|$ is a function about
$\eta_{n}$ which is finite.

To complement the theoretical analysis above, we further conduct an
empirical study to understand the actual distributions of $\hat{x}_{i}-x_{i}^{\mathrm{cvp}}$.
We enumerate the possible values of errors and their probabilities
using lattice reduction (using boosted LLL) aided ZF/SIC estimators
in Table \ref{tab:title1}. Note that these errors are not the decoding
error of VP, but they affect the decoding error of VP through resulted
SNR. Generally, more evident differences have smaller SNRs. In the
setup of the simulation, the probabilities are averaged from $10^{4}$
Monte Carlo runs, the size of constellations is set as $M=32$, and
the size of systems are set as $m=n=8,12,16,20$, respectively. Since
our simulations show that choosing other values of $M$ still yields
similar error distributions as in Table \ref{tab:title1}, we don't
present them in this paper. 

\begin{table*}[tbh]
\protect\protect\caption{The values $\hat{x}_{i}-x_{i}^{\mathrm{cvp}}$ with $i\in[n]$ and
their probabilities after ``Phase 1'' in hybrid precoding.}
\label{tab:title1} \centering %
\begin{tabular}{|c|c||c|c|c|c|c|c|c|c|c|}
\hline 
\multicolumn{1}{|c}{error} & distance & $-4$ & $-3$ & $-2$ & $-1$ & $0$ & $1$ & $2$ & $3$ & $4$\tabularnewline
\hline 
$n=8$ & LR-ZF & $0$ & $0$ & $0$ & $0.0666$ & $0.8670$ & $0.0664$ & $0$ & $0$ & $0$\tabularnewline
\cline{2-11} 
 & LR-SIC & $0$ & $0$ & $0.0001$ & $0.0505$ & $0.8973$ & $0.0518$ & $0.0001$ & $0$ & $0$\tabularnewline
\hline 
$n=12$ & LR-ZF & $0$ & $0$ & $0.0001$ & $0.0891$ & $0.8233$ & $0.0875$ & $0.0001$ & $0$ & $0$\tabularnewline
\cline{2-11} 
 & LR-SIC & $0$ & $0$ & $0.0013$ & $0.0817$ & $0.8348$ & $0.0808$ & $0.0013$ & $0$ & $0$\tabularnewline
\hline 
$n=16$ & LR-ZF & $0$ & $0$ & $0.0006$ & $0.1123$ & $0.7752$ & $0.1112$ & $0.0008$ & $0$ & $0$\tabularnewline
\cline{2-11} 
 & LR-SIC & $0$ & $0.0001$ & $0.0040$ & $0.1113$ & $0.7715$ & $0.1090$ & $0.0039$ & $0.0001$ & $0$\tabularnewline
\hline 
$n=20$ & LR-ZF & $0$ & $0.0001$ & $0.0022$ & $0.1342$ & $0.7284$ & $0.1327$ & $0.0024$ & $0.0001$ & $0$\tabularnewline
\cline{2-11} 
 & LR-SIC & $0.0001$ & $0.0007$ & $0.0082$ & $0.1348$ & $0.7119$ & $0.1352$ & $0.0086$ & $0.0005$ & $0$\tabularnewline
\hline 
\end{tabular}
\end{table*}

Table \ref{tab:title1} shows the values of error distance of both
LR-ZF and LR-SIC concentrate around $0$. It is clear that the range
of $\hat{x}_{i}-x_{i}^{\mathrm{cvp}}$ slowly grows w.r.t. the dimension
of the system; however, these values are much smaller than their theoretical
upper bounds. The statistical information provided by this empirical
study can be taken into account when designing threshold functions
for AMP.

\subsection{\label{sub:Prerequisites-for-AMP}Prerequisites for AMP}

Regarding the constellation of $\mathbf{x}$, we have demonstrated
that the error of ZF/SIC estimator is bounded to a function about
system dimension $n$ and some inherent lattice metrics. This means
we are not facing an infinite lattice decoding problem with $\mathbb{Z}$
constellations in Eq. (\ref{eq:AMP MODEL}), whence the application
of AMP becomes possible. Moreover, the bound of $\mathcal{B}^{n}$
can be made very small when designing our AMP algorithm.

Regarding the distribution of noise $\mathbf{w}^{\mathrm{amp}}=\mathbf{y}-\mathbf{H}\mathbf{x}$,
it is not known a priori. We can equip $\mathbf{w}^{\mathrm{amp}}$
with a Gaussian distribution $\mathrm{N}(\bm{0},\sigma^{2}\mathbf{I}_{m})$
with $0<\sigma^{2}<\left\Vert \mathbf{y}-\mathbf{H}\hat{\mathbf{x}}\right\Vert ^{2}/m$,
based on which we obtain the non-informative likelihood function of
$\mathbf{x}$:
\[
p(\mathbf{x})\sim\mathrm{N}(\mathbf{H}^{-1}\mathbf{y},\sigma^{2}(\mathbf{H}^{\top}\mathbf{H})^{-1}).
\]

Lastly, as for the channel matrix $\mathbf{H}$, if the basis now
has i.i.d. entries satisfying $\mathbb{E}H_{i,j}=0$ and $\mathbb{E}H_{i,j}^{2}=1/n$
and admitting sub-Gaussian tail conditions \cite{Bayati2015,Bayati2011},
which we refer to as sub-Gaussian conditions, then one can adopt the
well developed AMP \cite{Donoho2009,Montanari2010} or GAMP \cite{Rangan2014}
algorithms to solve our problem in Eq. (\ref{eq:AMP MODEL}) rigorously.
On the contrary, if a reduced basis is far from having sub-Gaussian
entries, then using AMP cannot provide any performance gain. Fortunately,
it is known that a basis is short and nearly orthogonal after lattice
reduction, which means its column-wise dependency is small. Moreover,
a reduced basis in VP often has ``small'' entries (in the sense
of \cite{Rangan2017}) such that the approximations in AMP are valid.
We further justified the two arguments above in Appendix \ref{sec:ODmu}.
As a result, we propose to describe a reduced basis with Gaussian
distributions and implement AMP on it, and the plausibility of this
method will be confirmed by simulations. Without loss of generality,
suppose $H_{i,j}\sim\mathrm{N}(0,\sigma_{j}^{2}/m)$ for $i\in[m]$,
then one can use the values of basis entries to obtain the maximum
likelihood estimation for each $\sigma_{j}^{2}$. To see this, note
that the likelihood function w.r.t. $\sigma_{j}^{2}$ and $m$ samples
$H_{1,j},\ldots,H_{m,j}$ is $L\left(H_{1,j},\ldots,H_{m,j},\sigma_{j}^{2}\right)=$
\begin{equation}
\frac{1}{\left(2\pi\sigma_{j}^{2}/m\right)^{n/2}}\exp\left(-\frac{1}{2\sigma_{j}^{2}/m}\sum_{i\in[m]}H_{i,j}^{2}\right),\label{eq:ml_esi_amp}
\end{equation}
 then it follows from solving $\partial L\left(H_{1,j},\ldots,H_{m,j},\sigma_{j}^{2}\right)/\partial\sigma_{j}^{2}=0$
that $\sigma_{j}^{2}=\sum_{i\in[m]}H_{i,j}^{2}$. Based on the above,
we will modify the AMP algorithm in \cite{Donoho2009,Montanari2010}
and analyze its performance in the next section.

\section{\label{sec:Our-AMP-L-algorithm} AMP algorithm for Eq. (\ref{eq:AMP MODEL})}

\begin{figure}[th]
\center

\includegraphics[clip,width=3.4in]{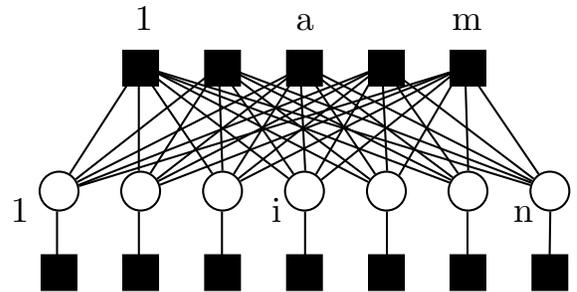}

\protect\caption{The factor graph associated to the probability distribution (\ref{eq:target map}).
Empty circles corresponds to variables, lines correspond to edges,
and solid squares correspond to factors. }
\label{fig_factor_graph} 
\end{figure}

Combing the non-informative likelihood function with the prior function
$p_{X}(x_{i})$, it yields a Maximum-a-Posteriori (MAP) function for
Bayesian estimation:
\begin{equation}
p(\mathbf{x}|\mathbf{y},\mathbf{H})\varpropto\prod_{a\in[m]}p_{a}(\mathbf{x},y_{a})\prod_{i\in[n]}p_{X}(x_{i}),\label{eq:target map}
\end{equation}
 where $p_{a}(\mathbf{x},y_{a})=\exp(-\frac{1}{2\sigma^{2}}(y_{a}-\mathbf{H}_{a,1:n}\mathbf{x})^{2})$,
and $p_{X}(x_{i})$ will be designed in Section \ref{sub:sparse_prior}.
The factorized structure in (\ref{eq:target map}) can be conveniently
described by a factor graph \cite{Kschischang2001,Bera2017}. It includes
a variable node for each $x_{i}$, a factor node for each $p_{X}(x_{i})$,
and a factor node for each $p_{a}(\mathbf{x},y_{a})$, where $i\in[n]$,
$a\in[m]$. If $x_{i}$ appears in $p_{X}(x_{i})$ or $p_{a}(\mathbf{x},y_{a})$,
then they are connected by an edge. Clearly, $x_{i}$ and $p_{a}(\mathbf{x},y_{a})$
are connected if and only if and only if $H_{a,i}\neq0$. Such a factor
graph is reproduced in Fig. \ref{fig_factor_graph}. 

In the sequel, we first show how to simplify BP to reach an AMP algorithm
by using the approximation techniques in \cite{Donoho2009,Montanari2010,Maleki2011}.
After that, we will characterize the symbol-wise estimation errors
in Theorem \ref{thm:SE non-uniform power} and present the threshold
functions of certain prior distributions.

\subsection{\label{sub:Approximating-BP}Simplified BP}

Each factor graph naturally induces a BP algorithm \cite{Donoho2009a}
that involves two types of messages: messages from variable nodes
to factor nodes denoted by $J_{i\rightarrow a}(x_{i})$, and messages
from factor nodes to variable nodes, denoted by $\hat{J}_{a\rightarrow i}(x_{i})$.
Here, messages refer to probability distribution functions, which
are recursively updated to compute marginal posterior density functions
for the variables. At the $t$th iteration, they are updated as follows
\begin{equation}
\hat{J}_{a\rightarrow i}^{t}(x_{i})=\int_{\mathbf{x}\backslash x_{i}}\{p_{a}(\mathbf{x},y_{a})\prod_{j\in[n]\backslash i}J_{j\rightarrow a}^{t}(x_{j})\}\mathrm{d}\mathbf{x},\label{EqMessages2}
\end{equation}

\begin{equation}
J_{i\rightarrow a}^{t+1}(x_{i})=p_{X}(x_{i})\prod_{b\in[m]\backslash a}\hat{J}_{b\rightarrow i}^{t}(x_{i}).\label{EqMessages1}
\end{equation}

These messages are impractical to evaluate in the Lebesgue measure
space, and thus often simplified by approximation techniques. We make
the approximation from an expectation propagation \cite{Minka2001}
perspective hereby. Suppose the message in Eq. (\ref{EqMessages2})
is estimated by a Gaussian function with mean $\alpha_{a\rightarrow i}^{t}/\beta_{a\rightarrow i}^{t}$
and variance $1/\beta_{a\rightarrow i}^{t}$, then 
\begin{eqnarray}
\hat{J}_{a\rightarrow i}^{t}(x_{i}) & = & \mathrm{N}(H_{a,i}x_{i};\alpha_{a\rightarrow i}^{t}/\beta_{a\rightarrow i}^{t},1/\beta_{a\rightarrow i}^{t}).\label{eq:factor2vari}
\end{eqnarray}
By substituting Eq. (\ref{eq:factor2vari}) into Eq. (\ref{EqMessages1}),
we have 
\begin{align}
J_{i\rightarrow a}^{t+1}(x_{i}) & \varpropto p_{X}(x_{i})\exp((\sum_{b\in[m]\backslash a}H_{b,i}\alpha_{b\rightarrow i}^{t})x_{i}\nonumber \\
 & -1/2(\sum_{b\in[m]\backslash a}H_{b,i}^{2}\beta_{b\rightarrow i}^{t})x_{i}^{2}+O(nH_{a,i}^{3}x_{i}^{3}))\nonumber \\
 & \varpropto p_{X}(x_{i})\mathrm{N}(x_{i};u_{i\rightarrow a},v_{i\rightarrow a}),\label{eq:i2a2}
\end{align}
 where 
\begin{equation}
u_{i\rightarrow a}^{t}=\frac{\sum_{b\in[m]\backslash a}H_{b,i}\alpha_{b\rightarrow i}^{t}}{\sum_{b\in[m]\backslash a}H_{b,i}^{2}\beta_{b\rightarrow i}^{t}},\label{eq:u}
\end{equation}
\begin{equation}
v_{i\rightarrow a}^{t}=\frac{1}{\sum_{b\in[m]\backslash a}H_{b,i}^{2}\beta_{b\rightarrow i}^{t}}.\label{eq:v}
\end{equation}
In the other direction, we work out messages $J_{i\rightarrow a}^{t+1}(x_{i})$
with Gaussian functions through matching their first and second order
moments by the following constraints:

\begin{equation}
J_{i\rightarrow a}^{t+1}(x_{i})=\mathrm{N}(x_{i};\eta(u_{i\rightarrow a}^{t},v_{i\rightarrow a}^{t}),\mathrm{\kappa}(u_{i\rightarrow a}^{t},v_{i\rightarrow a}^{t})),\label{eq:mes va2fa}
\end{equation}

\begin{equation}
\mathrm{\eta}(u_{i\rightarrow a}^{t},v_{i\rightarrow a}^{t})=\int_{x}xp_{X}(x)\mathrm{N}(x;\thinspace u_{i\rightarrow a}^{t},v_{i\rightarrow a}^{t})\mathrm{d}x,\label{eq:eta first}
\end{equation}

\begin{align}
\mathrm{\kappa}(u_{i\rightarrow a}^{t},v_{i\rightarrow a}^{t}) & =\int_{x}x^{2}p_{X}(x)\mathrm{N}(x;\thinspace u_{i\rightarrow a}^{t},v_{i\rightarrow a}^{t})\mathrm{d}x\nonumber \\
 & -\mathrm{\eta}^{2}(u_{i\rightarrow a}^{t},v_{i\rightarrow a}^{t}),\label{eq:kappa first}
\end{align}
where $\mathrm{\eta}(u_{i\rightarrow a}^{t},v_{i\rightarrow a}^{t})$
and $\mathrm{\kappa}(u_{i\rightarrow a}^{t},v_{i\rightarrow a}^{t})$
are posterior mean and variance functions, respectively. We will refer
to them as threshold functions. From Eq. (\ref{eq:mes va2fa}), inferring
$x_{i\rightarrow a}^{t+1}$ and its variance $\varsigma_{i\rightarrow a}^{t+1}$
from $J_{i\rightarrow a}^{t+1}(x_{i})$ by using the MAP principle
yields: 
\begin{equation}
x_{i\rightarrow a}^{t+1}=\eta(u_{i\rightarrow a}^{t},v_{i\rightarrow a}^{t}),\label{eq:xhat}
\end{equation}
\begin{equation}
\varsigma_{i\rightarrow a}^{t+1}=\mathrm{\kappa}(u_{i\rightarrow a}^{t},v_{i\rightarrow a}^{t}).\label{eq:vhat}
\end{equation}
By substituting the approximation of Eq. (\ref{eq:mes va2fa}) into
Eq. (\ref{EqMessages2}), which becomes a multidimensional Gaussian
function expectation $\mathbb{E}(p_{a}(\mathbf{x},y_{a}))$ w.r.t.
probability measure $\prod_{j\in[n]\backslash i}J_{j\rightarrow a}^{t}(x_{j})$,
the integration over Gaussian functions becomes $\hat{J}_{a\rightarrow i}^{t}(x_{i})\varpropto$
\begin{equation}
\mathrm{N}(H_{a,i}x_{i};\thinspace y_{a}-\sum_{j\in[n]\backslash i}H_{a,j}x_{j\rightarrow a}^{t-1},\sigma^{2}+\sum_{j\in[n]\backslash i}|H_{a,j}|^{2}\varsigma_{j\rightarrow a}^{t-1}).\label{eq:cmplast}
\end{equation}

Compare Eq. (\ref{eq:cmplast}) with the previously defined mean $\alpha_{a\rightarrow i}^{t}/\beta_{a\rightarrow i}^{t}$
and variance $1/\beta_{a\rightarrow i}^{t}$, we have

\begin{equation}
\alpha_{a\rightarrow i}^{t}=(y_{a}-\sum_{j\in[n]\backslash i}H_{a,j}x_{j\rightarrow a}^{t-1})/(\sigma^{2}+\sum_{j\in[n]\backslash i}|H_{a,j}|^{2}\varsigma_{j\rightarrow a}^{t-1}),\label{eq:r}
\end{equation}
 
\begin{equation}
\beta_{a\rightarrow i}^{t}=1/(\sigma^{2}+\sum_{j\in[n]\backslash i}|H_{a,j}|^{2}\varsigma_{j\rightarrow a}^{t-1}).\label{eq:beta}
\end{equation}

Thus far, Eqs . (\ref{eq:u}) (\ref{eq:v}) (\ref{eq:xhat}) (\ref{eq:vhat})
(\ref{eq:r}) (\ref{eq:beta}) define a simplified version of BP,
where the tracking of $2mn$ functions in Eqs. (\ref{EqMessages1})
and (\ref{EqMessages2}) has been replaced by the tracking of $6mn$
scalars.
\begin{rem}
Our derivation is to equip $\hat{J}_{a\rightarrow i}^{t}(x_{i})$
with a density function that can be fully described by its first and
second moments, then one obtains their moment equations when passing
$J_{j\rightarrow a}^{t}(x_{j})$ back. In \cite[Lem. 5.3.1]{Maleki2011},
Maleki had applied the Berry\textendash Esseen theorem to prove that
approximating $\hat{J}_{a\rightarrow i}^{t}(x_{i})$ with a Gaussian
is tight. Although our variance $1/\beta_{a\rightarrow i}^{t}$ of
$\hat{J}_{a\rightarrow i}^{t}(x_{i})$ looks different from his, they
are indeed equivalent if we set the variance $\varsigma_{i\rightarrow a}^{t}$
of $J_{i\rightarrow a}^{t}(x_{i})$ as $\sigma^{2}\varsigma_{i\rightarrow a}^{t}$.
Moreover, \cite[Lem. 5.5.4]{Maleki2011} also justifies the correctness
on the other side of our approximation.
\end{rem}

\subsection{Reaching $O(m+n)$ scalars}

For a reduced lattice basis $\mathbf{H}$, recall that we have set
$\sigma_{1}^{2}=\left\Vert \mathbf{h}_{1}\right\Vert ^{2},\ldots\thinspace,\sigma_{n}^{2}=\left\Vert \mathbf{h}_{n}\right\Vert ^{2}$,
and the statistical variance for each entry of $\mathbf{H}$ is $\mathbb{V}(H_{b,i})=\sigma_{i}^{2}/m$.
Then we can employ this knowledge to further simplify the algorithm
in Section \ref{sub:Approximating-BP}. Here we define 
\begin{equation}
r_{a\rightarrow i}^{t}=\alpha_{a\rightarrow i}^{t}/\beta_{a\rightarrow i}^{t}=y_{a}-\sum_{j\in[n]\backslash i}H_{a,j}x_{j\rightarrow a}^{t-1}.\label{eq:second-r}
\end{equation}
By equipping all the $\beta_{b\rightarrow i}^{t}$ with equal magnitude,
referred to as $\beta_{\bar{b}\rightarrow i}^{t}$, as well as using
$\sum_{b\in[m]\backslash a}H_{b,i}^{2}\approx\sigma_{i}^{2}$ due
to the law of large numbers, it yields 
\begin{equation}
x_{i\rightarrow a}^{t}=\eta(\frac{1}{\sigma_{i}^{2}}\sum_{b\in[m]\backslash a}H_{b,i}r_{b\rightarrow i}^{t},\frac{1}{\sigma_{i}^{2}\beta_{\bar{b}\rightarrow i}^{t}}),\label{eq:second-x}
\end{equation}

\begin{equation}
\varsigma_{i\rightarrow a}^{t}=\mathrm{\kappa}(\frac{1}{\sigma_{i}^{2}}\sum_{b\in[m]\backslash a}H_{b,i}r_{b\rightarrow i}^{t},\frac{1}{\sigma_{i}^{2}\beta_{\bar{b}\rightarrow i}^{t}}).\label{eq:second-x-1}
\end{equation}
For the moment, we can expand the local estimations about $r_{a\rightarrow i}^{t}$
and $x_{i\rightarrow a}^{t}$ as $r_{a\rightarrow i}^{t}=r_{a}^{t}+\delta r_{a\rightarrow i}^{t}$,
$x_{i\rightarrow a}^{t}=x_{i}^{t}+\delta x_{i\rightarrow a}^{t}$,
so the techniques in \cite{Montanari2010,Donoho2009a} can be employed.
The crux of these transformation is to neglect elements whose amplitudes
are no larger than $O(1/n)$. Subsequently, Eqs. (\ref{eq:second-r})
and (\ref{eq:second-x}) become
\begin{equation}
r_{a}^{t}+\delta r_{a\rightarrow i}^{t}=y_{a}-\sum_{j\in[n]}H_{a,j}(x_{j}^{t-1}+\delta x_{j\rightarrow a}^{t-1})+H_{a,i}x_{i}^{t-1},\label{eq:deltar1}
\end{equation}

\begin{equation}
x_{i}^{t}+\delta x_{i\rightarrow a}^{t}=\eta(\frac{1}{\sigma_{i}^{2}}\sum_{b\in[m]}H_{b,i}(r_{b}^{t}+\delta r_{b\rightarrow i}^{t})-\frac{1}{\sigma_{i}^{2}}H_{a,i}r_{a}^{t},\thinspace\frac{1}{\sigma_{i}^{2}\beta_{\bar{b}\rightarrow i}^{t}}).\label{eq:deltax1}
\end{equation}
 In (\ref{eq:deltar1}), terms with common $\{i\}$ indexes are mutually
related while others are not, so that
\begin{equation}
r_{a}^{t}=y_{a}-\sum_{j\in[n]}H_{a,j}(x_{j}^{t-1}+\delta x_{j\rightarrow a}^{t-1}),\label{eq:r20}
\end{equation}

\begin{equation}
\delta r_{a\rightarrow i}^{t}=H_{a,i}x_{i}^{t-1}.\label{eq:r21}
\end{equation}
 Further expand the r.h.s. of (\ref{eq:deltax1}) with the first order
Taylor expression of $\eta(u,v)$ at $u$, in which
\begin{gather}
\frac{\partial\eta(u,v)}{\partial u}\mid_{u=\frac{1}{\sigma_{i}^{2}}\sum_{b\in[m]\backslash a}H_{b,i}r_{b\rightarrow i}^{t},v=\frac{1}{\sigma_{i}^{2}\beta_{\bar{b}\rightarrow i}^{t}}}=\nonumber \\
\sigma_{i}^{2}\beta_{\bar{b}\rightarrow i}^{t}\mathrm{\kappa}(\frac{1}{\sigma_{i}^{2}}\sum_{b\in[m]\backslash a}H_{b,i}r_{b\rightarrow i}^{t},\frac{1}{\sigma_{i}^{2}\beta_{\bar{b}\rightarrow i}^{t}}),
\end{gather}
then it yields

\begin{gather*}
x_{i}^{t}+\delta x_{i\rightarrow a}^{t}=\eta(\frac{1}{\sigma_{i}^{2}}\sum_{b\in[m]}H_{b,i}(r_{b}^{t}+\delta r_{b\rightarrow i}^{t}),\frac{1}{\sigma_{i}^{2}\beta_{\bar{b}\rightarrow i}^{t}})\\
-\beta_{\bar{b}\rightarrow i}^{t}\mathrm{\kappa}(\frac{1}{\sigma_{i}^{2}}\sum_{b\in[m]}H_{b,i}(r_{b}^{t}+\delta r_{b\rightarrow i}^{t}),\frac{1}{\sigma_{i}^{2}\beta_{\bar{b}\rightarrow i}^{t}})H_{a,i}r_{a}^{t}.
\end{gather*}
Distinguishing terms that are dependent on indexes $\left\{ a\right\} $
leads to
\begin{equation}
x_{i}^{t}=\eta(\frac{1}{\sigma_{i}^{2}}\sum_{b\in[m]}H_{b,i}(r_{b}^{t}+\delta r_{b\rightarrow i}^{t}),\frac{1}{\sigma_{i}^{2}\beta_{\bar{b}\rightarrow i}^{t}}),\label{eq:x22}
\end{equation}

\begin{equation}
\delta x_{i\rightarrow a}^{t}=-\beta_{\bar{b}\rightarrow i}^{t}\mathrm{\kappa}(\frac{1}{\sigma_{i}^{2}}\sum_{b\in[m]}H_{b,i}(r_{b}^{t}+\delta r_{b\rightarrow i}^{t}),\frac{1}{\sigma_{i}^{2}\beta_{\bar{b}\rightarrow i}^{t}})H_{a,i}r_{a}^{t}.\label{eq:x23}
\end{equation}
 Then we substitute (\ref{eq:r21}) into (\ref{eq:x22}), and (\ref{eq:x23})
into (\ref{eq:r20}), to obtain
\begin{equation}
x_{i}^{t}=\eta(\frac{1}{\sigma_{i}^{2}}\sum_{b\in[m]}H_{b,i}r_{b}^{t}+x_{i}^{t-1},\frac{1}{\sigma_{i}^{2}\beta_{\bar{b}\rightarrow i}^{t}}),\label{eq:25}
\end{equation}
\begin{equation}
r_{a}^{t}=y_{a}-\sum_{j\in[n]}H_{a,j}x_{j}^{t-1}+\phi r_{a}^{t-1},\label{eq:26}
\end{equation}
 where 
\begin{equation}
\phi=\sum_{j\in[n]}H_{a,j}^{2}\beta_{\bar{b}\rightarrow j}^{t-1}\mathrm{\kappa}(\frac{1}{\sigma_{i}^{2}}\sum_{b\in[m]}H_{b,i}(r_{b}^{t}+\delta r_{b\rightarrow i}^{t}),\frac{1}{\sigma_{i}^{2}\beta_{\bar{b}\rightarrow i}^{t}}).\label{eq:27}
\end{equation}

\subsection{Further simplification}

From (\ref{eq:25}), the estimated variance for each $x_{i}^{t}$
now becomes
\begin{equation}
\varsigma_{i}^{t}=\kappa(\frac{1}{\sigma_{i}^{2}}\sum_{b\in[m]}H_{b,i}r_{b}^{t}+x_{i}^{t-1},\frac{1}{\sigma_{i}^{2}\beta_{\bar{b}\rightarrow i}^{t}}),\label{eq:}
\end{equation}
As $\varsigma_{i}^{t}\thickapprox\varsigma_{i\rightarrow b}^{t},\thinspace\forall\thinspace b$,
(\ref{eq:beta}) tells 
\begin{equation}
\beta_{\bar{b}\rightarrow i}^{t}=1/(\sigma^{2}+\frac{\sum_{j\in[n]}\sigma_{j}^{2}\varsigma_{j}^{t-1}}{m}).\label{eq:beta-1}
\end{equation}
 According to (\ref{eq:beta-1}), we denote $\beta_{\bar{b}\rightarrow i}^{t}$
as $1/\tau_{t}^{2}$, then the whole algorithm can be described by
the following four steps: 
\begin{equation}
x_{i}^{t}=\eta(1/\sigma_{i}^{2}\sum_{b\in[m]}H_{b,i}r_{b}^{t}+x_{i}^{t-1},\tau_{t}^{2}/\sigma_{i}^{2}),\label{eq:-1}
\end{equation}

\begin{equation}
\varsigma_{i}^{t}=\kappa(1/\sigma_{i}^{2}\sum_{b\in[m]}H_{b,i}r_{b}^{t}+x_{i}^{t-1},\tau_{t}^{2}/\sigma_{i}^{2}),\label{eq:-2}
\end{equation}

\begin{equation}
r_{a}^{t+1}=y_{a}-\sum_{j\in[n]}H_{a,j}x_{j}^{t}+\frac{\sum_{j\in[n]}\sigma_{j}^{2}\varsigma_{j}^{t}}{m\tau_{t}^{2}}r_{a}^{t},\label{eq:26-1}
\end{equation}

\begin{equation}
\tau_{t+1}^{2}=\sigma^{2}+\frac{\sum_{j\in[n]}\sigma_{j}^{2}\varsigma_{j}^{t}}{m}.\label{eq:beta-1-1}
\end{equation}
Denote $\bar{\tau}_{t}^{2}=1/n\sum_{j\in[n]}\sigma_{j}^{2}\varsigma_{j}^{t}$
, then iterations in (\ref{eq:-1}) to (\ref{eq:beta-1-1}) can be
compactly represented by matrix-vector products. 

Further incorporate some implementation details, our AMP algorithm
is summarized in Algorithm \ref{AlgoAMP}. 

\begin{algorithm}[ht!] \label{AlgoAMP} \caption{The AMP  algorithm.} 
	 \KwIn{Lattice basis $ \mathbf{H}=[\mathbf{h}_{1},\ldots\thinspace,\mathbf{h}_{n}]$, target $\mathbf{y}$, number of iterations $T$, threshold functions $\eta$ and $\kappa$, variance parameter $\sigma^2$.}  \KwOut{estimated coefficient vector $\mathbf{x}^{\mathrm{amp}}$.}   $\mathbf{x}^{0}=\mathbf{0}$,  $f_0=\left\Vert \mathbf{y}\right\Vert ^{2}$,  $\mathbf{r}^{1}=\mathbf{y}$, ${\tau}_{1}^{2}=10^4$\;   \For{$i=1,\ldots,n$}{$\sigma_{i}^{2}=\left\Vert \mathbf{h}_{i}\right\Vert ^{2}$}  $\Theta=\mathrm{diag}\left(1/\sigma_{1}^{2},\ldots,1/\sigma_{n}^{2}\right)$\;   \For{$t=1,\ldots, T$}    {  $\mathbf{x}^{t}=\eta(\Theta\mathbf{H}^{\top}\mathbf{r}^{t}+  \mathbf{x}^{t-1},\Theta\tau_{t}^{2}\mathbf{1})$\;  $\bar{\tau}_{t}^{2}=\langle\Theta^{-1}\kappa(\Theta\mathbf{H}^{\top}\mathbf{r}^{t}+\mathbf{x}^{t-1},\Theta\tau_{t}^{2}\mathbf{1})\rangle$\; $\mathbf{r}^{t+1}=\mathbf{y}-\mathbf{H}\mathbf{x}^{t}+\frac{n}{m}\frac{\bar{\tau}_{t}^{2}}{\tau_{t}^{2}}\mathbf{r}^{t}$\; $\tau_{t+1}^{2}=\sigma^{2}+\frac{n}{m}\bar{\tau}_{t}^{2}$ \;  	$f_i=\left\Vert \mathbf{y}-\mathbf{H} \lfloor \mathbf{x}^{t}\rceil \right\Vert ^{2} $ ;   \Comment{Record the fitness values}\;  }   	$i'=\arg\min_{i} f_i$\; $\mathbf{x}^{\mathrm{amp}}=\lfloor \mathbf{x}^{i'}\rceil$.  \end{algorithm}

\subsection{Performance and Discussions}

One advantage of using AMP is that we can exactly analyze the mean
square errors of the estimation, as shown in the following theorem.
Its proof is given in Appendix \ref{sec:Proof-of-se}.
\begin{thm}
\label{thm:SE non-uniform power} Let the reduced lattice basis be
modeled as $H_{b,i}\sim\mathrm{N}(0,\sigma_{i}^{2}/m)$, with $b\in[m]$,
$i\in[n]$, and denote $\bar{\mathbf{x}}=\mathbf{x}^{\mathrm{cvp}}-\hat{\mathbf{x}}$
as the desired estimation. For each $\mathbf{x}^{t}$ provided by
Algorithm \ref{AlgoAMP}, as $n$ goes to infinity and $m$ grows
in the same order as $n$, we have almost surely for all $i$ that:
\[
\left\Vert x_{i}^{t}-\bar{x}_{i}\right\Vert ^{2}=\mathbb{E}|\eta(X+\tau_{t,i}Z,\tau_{t,i}^{2})-X|^{2},
\]
where $\tau_{t,i}$ admits the following iteration relation: 
\begin{equation}
\tau_{t,i}^{2}=\frac{1}{m\sigma_{i}^{2}}\sum_{j\in[n]}\sigma_{j}^{2}\mathbb{E}|\eta(X+\tau_{(t-1),j}Z,\tau_{(t-1),j}^{2})-X|^{2}+\frac{\sigma^{2}}{\sigma_{i}^{2}},\label{eq:se standard}
\end{equation}
and the expectation is taken over two independent random variables
$Z\sim\mathrm{N}(0,1)$ and $X\sim p_{X}$. 
\end{thm}
By defining $\widetilde{\tau}_{t}^{2}\triangleq\tau_{t,j}^{2}\sigma_{j}^{2}=\tau_{t,i}^{2}\sigma_{i}^{2}$,
Eq. (\ref{eq:se standard}) becomes
\begin{equation}
\widetilde{\tau}_{t}^{2}=\frac{1}{m}\sum_{j\in[n]}\sigma_{j}^{2}\mathbb{E}|\eta(X+\widetilde{\tau}_{t-1}/\sigma_{j}Z,\widetilde{\tau}_{t-1}^{2}/\sigma_{j}^{2})-X|^{2}+\sigma^{2}.\label{eq:alSE}
\end{equation}
 The above equation is referred to as the state evolution equation
for our AMP. Based on this equation, we will study the impact of parameters
in the threshold functions.

Although one may recognize that the AMP/GAMP algorithms in \cite{Jeon2015,Donoho2009,Rangan2011}
may also be employed for our ``Phase 2'' estimation after further
regularizing the channels (i.e., let $\mathbf{H}\leftarrow\mathbf{H}\Theta^{1/2}$
and consider $\mathbf{x}\leftarrow\Theta^{-1/2}\mathbf{x}$), the
derived AMP can provide the following valuable insights: i) We can
explicitly study the impact of channel powers $\sigma_{i}^{2}$'s
on the state evolution equation based on our derivation (e.g., Proposition
\ref{prop:powerNepsi}). All the $\sigma_{i}^{2}$'s are obtained
by using the maximum likelihood estimator (\ref{eq:ml_esi_amp}).
ii) The estimated data symbols in Algorithm \ref{AlgoAMP} is reflecting
the MAP estimation without the need of further regularization.

\section{\label{sub:sparse_prior}Designing threshold functions}

The AMP algorithm needs to work with certain threshold functions which
are designed according to specific, definite information about coefficient
vector $\mathbf{x}$. It is noteworthy that the theoretical bounds
of $\mathcal{B}^{n}$ are derived from a worst case analysis which
are often very large. However, we don't need to adopt these bounds
for designing threshold functions due to the following two reasons.
First, LR aided ZF/SIC are in practice quite close to sphere decoding
in small dimensions and there also exist certain probabilities that
the error distance is small in large dimensions, so it suffices to
impose a ternary distribution for these scenarios. Second, although
we recognize that $\max_{i}|\hat{x}_{i}-x_{i}^{\mathrm{cvp}}|$ would
increase as the system dimension grows, where $\hat{x}_{i}-x_{i}^{\mathrm{cvp}}$
admits a distribution in the shape of a discrete Gaussian, a threshold
function based on this distribution is not only numerically unstable
\cite[P. 182]{Jeon2016} but also requires the basis matrix to be
extremely tall \cite{Jeon2015}. Therefore, an efficient way to use
 such discrete prior knowledge is to use linear estimation based on
continuous Gaussian distributions \cite{lyu15,Jeon2016}.

In this section, we present threshold functions for a ternary distribution
and a discrete Gaussian distribution.

\subsection{Ternary Distribution}

According to the empirical study above, a dominant portion of ``errors''
could be corrected by only imposing a ternary distribution $\left\{ -1,0,1\right\} $
for $p_{X}(x_{i})$. Here, we present its threshold functions $\eta_{\varepsilon}(u,v)$
and $\kappa_{\varepsilon}(u,v)$ in the following lemma.
\begin{lem}
\label{lem:sparseFilter}Let $Y=X+W$, with $X\sim p_{X}(x)=(1-\varepsilon)\delta(x)+\varepsilon/2\delta(x-1)+\varepsilon/2\delta(x+1)$,
$W\sim\mathrm{N}(0,v)$. Then the conditional mean and conditional
variance of $X$ on $Y$ are: 
\begin{equation}
\eta_{\varepsilon}(u,v)\triangleq\mathbb{E}(X|Y=u)=\frac{\sinh(u/v)}{(1-\varepsilon)/\varepsilon e^{1/(2v)}+\cosh(u/v)},\label{eq:SPAeta}
\end{equation}
\begin{equation}
\kappa_{\varepsilon}(u,v)\triangleq\mathbb{V}(X|Y=u)=\frac{(1-\varepsilon)/\varepsilon e^{1/(2v)}\cosh(u/v)+1}{((1-\varepsilon)/\varepsilon e^{1/(2v)}+\cosh(u/v))^{2}}.\label{eq:spakapa}
\end{equation}

\end{lem}

\begin{figure}[th]
\center

\includegraphics[clip,width=3.4in]{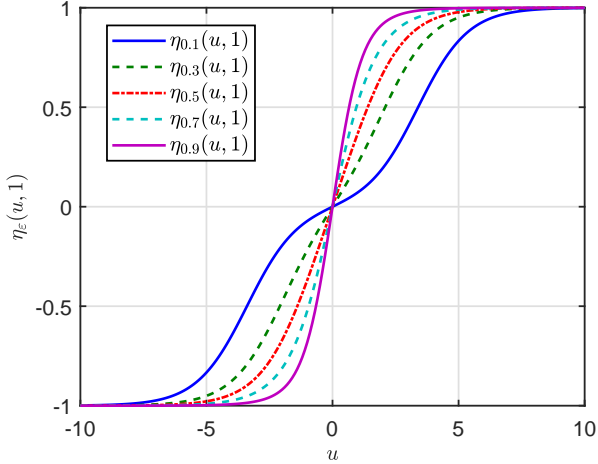}

\protect\protect\caption{The threshold function $\eta_{\varepsilon}(u,v)$.}
\label{fig FILTER} 
\end{figure}

\begin{IEEEproof}
Since the posterior probability is proportional to the likelihood
multiplied by the prior probability, we have
\begin{align*}
 & P_{X\mid Y=u}\left(x\right)\\
 & \propto P_{X}\left(x\right)P_{Y=u\mid X}\left(y\right)\\
 & \propto\left[(1-\varepsilon)\delta(x)+\frac{\varepsilon}{2}\delta(x-1)+\frac{\varepsilon}{2}\delta(x+1)\right]\exp\left(-\frac{\left(x-u\right)^{2}}{2v}\right)\\
 & =\begin{cases}
(1-\varepsilon)\left(-\frac{u^{2}}{2v}\right)/S, & x=0,\\
\frac{\varepsilon}{2}\left(-\frac{\left(u-1\right)^{2}}{2v}\right)/S, & x=1,\\
\frac{\varepsilon}{2}\left(-\frac{\left(u+1\right)^{2}}{2v}\right)/S, & x=-1,
\end{cases}
\end{align*}
 where $S=(1-\varepsilon)\left(-\frac{u^{2}}{2v}\right)+\frac{\varepsilon}{2}\left(-\frac{\left(u-1\right)^{2}}{2v}\right)+\frac{\varepsilon}{2}\left(-\frac{\left(u+1\right)^{2}}{2v}\right)$.
Therefore, the conditional mean is
\[
\sum_{x}xP_{X\mid Y=u}\left(x\right)=\frac{\sinh(u/v)}{(1-\varepsilon)/\varepsilon e^{1/(2v)}+\cosh(u/v)},
\]
 and the conditional variance is 
\begin{align*}
 & \sum_{x}\left(x-\mathbb{E}(X|Y=u)\right)^{2}P_{X\mid Y=u}\left(x\right)\\
 & =\frac{(1-\varepsilon)/\varepsilon e^{1/(2v)}\cosh(u/v)+1}{((1-\varepsilon)/\varepsilon e^{1/(2v)}+\cosh(u/v))^{2}}.
\end{align*}

\end{IEEEproof}
These threshold functions have closed forms and are easy to compute.
The AMP algorithm using (\ref{eq:SPAeta}) (\ref{eq:spakapa}) from
a \textbf{t}ernary distribution is referred to as AMPT. In Fig. \ref{fig FILTER},
we have plotted function $\eta_{\varepsilon}(u,v)$ by setting $v=1$
and $\varepsilon\in\left\{ 0.1,0.3,0.5,0.7,0.9\right\} $.

\subsection{\label{sub:gau_prior-1}Gaussian Distribution}

The aim of this section is to explain how to obtain a closed-form
expression for threshold functions targeting discrete Gaussian distributions.
A discrete Gaussian distribution over $\mathbb{Z}$ with zero mean
and width $\sigma_{g}$ is defined as 
\[
\rho_{\sigma_{g}}(z)=\frac{1}{S}e^{-z^{2}/\left(2\sigma_{g}^{2}\right)},
\]
where $S=\sum_{k=-\infty}^{\infty}e^{-k^{2}/\left(2\sigma_{g}^{2}\right)}$.
According to a tail bound on discrete Gaussian \cite[Lem. 4.4]{Lyubashevsky2012},
we have 
\[
\mathrm{Pr}_{z\sim\rho_{\sigma_{g}}(z)}\left(|z|>k\sigma_{g}\right)\leq2e^{-k^{2}/2}
\]
 for any $k>0$. This implies that $\rho_{\sigma_{g}}(z)$ can be
calculated from a finite range. E.g., we have $\mathrm{Pr}_{z\sim\rho_{\sigma_{g}}(z)}\left(|z|>10\sigma_{g}\right)\leq3.86\times10^{-22}$.
If $\sigma_{g}=0.1$, then $\rho_{\sigma_{g}}(z)$ becomes equivalent
to the ternary distribution with $\varepsilon\leq0.5$.

Assume that we have observed $Y=u$ from model $Y=X+W$, with $X\sim p_{X}(x)=\rho_{\sigma_{g}}(x)$,
$W\sim\mathrm{N}(0,v)$. Then the threshold functions are given by
\begin{align*}
\eta_{d}(u,v) & =\frac{1}{S_{k}}\sum_{l=-k}^{k}le^{-\frac{l^{2}}{2\sigma_{g}^{2}}-\frac{\left(l-u\right)^{2}}{2v}},\\
\kappa_{d}(u,v) & =\frac{1}{S_{k}}\sum_{l=-k}^{k}\left(l-\eta_{g}(u,v)\right)^{2}e^{-\frac{l^{2}}{2\sigma_{g}^{2}}-\frac{\left(l-u\right)^{2}}{2v}},
\end{align*}
where $S_{k}=\sum_{l=-k}^{k}e^{-l^{2}/2\sigma_{g}^{2}-\left(l-u\right)^{2}/\left(2v\right)}$.
Recall that we have mentioned evaluating $\eta_{d}(u,v)$ and $\kappa_{d}(u,v)$
is generally computationally intensive, and the fixed points of their
state evolution equation are unfathomable. Fortunately, the sum of
a discrete Gaussian and a continuous Gaussian resembles a continuous
Gaussian if the discrete Gaussian is smooth \cite[Lem. 9]{Ling2014},
so we can replace $\rho_{\sigma_{g}}(x)$ with $\mathrm{N}(x;0,\sigma_{g}^{2})$
with properly chosen $\sigma_{g}^{2}$. Let the signal distribution
be $p_{X}(x)=\mathrm{N}(x;0,\sigma_{g}^{2})$, then it corresponds
to another pair of threshold functions that have closed-forms: 
\begin{align}
\eta_{g}(u,v) & =\frac{u\sigma_{g}^{2}}{\sigma_{g}^{2}+v},\label{eq:-3}\\
\kappa_{g}(u,v) & =\frac{v\sigma_{g}^{2}}{\sigma_{g}^{2}+v}.\label{eq:-4}
\end{align}
The AMP algorithm using (\ref{eq:-3}) (\ref{eq:-4}) due to \textbf{G}aussian
distributions is referred to as AMPG.

\subsection{\label{sub:The-impact-of_power}Parameters in Threshold Functions}

In this section, we will inspect the effect of chosen parameters on
the AMP algorithm, where the technique involved is about analyzing
fixed points (see \cite{Zheng2017tit} for more backgrounds). First,
the state evolution equation without the iteration subscript reads
\begin{equation}
\Psi(\widetilde{\tau}^{2})\triangleq\frac{1}{m}\sum_{j\in[n]}\sigma_{j}^{2}\mathbb{E}|\eta(X+\widetilde{\tau}/\sigma_{j}Z,\widetilde{\tau}^{2}/\sigma_{j}^{2})-X|^{2}+\sigma^{2}.\label{eq:general thres}
\end{equation}
We refer to $\widetilde{\tau}^{2}$ as a fixed point of $\Psi(\widetilde{\tau}^{2})$
if $\Psi(\widetilde{\tau}^{2})=\widetilde{\tau}^{2}$. A fixed point
is called stable if there exists $\epsilon\rightarrow0^{+}$, such
that $\Psi(\widetilde{\tau}^{2}+\epsilon)<\widetilde{\tau}^{2}$ and
$\Psi(\widetilde{\tau}^{2}-\epsilon)>\widetilde{\tau}^{2}$. When
$\Psi(0)=0$, the stability condition is relaxed to $\Psi(\widetilde{\tau}^{2}+\epsilon)<\widetilde{\tau}^{2}$.
A fixed point is called unstable if it fails the stability condition.
The estimation error of AMP is the smallest (resp. largest) if its
$\Psi(\widetilde{\tau}^{2})$ converges to the lowest (resp. highest)
stable fixed points \cite{Zheng2017tit}. 

For AMPT, we can demonstrate the impact of channel power $\left\{ \sigma_{j}^{2}\right\} $
and sparsity $\left(1-\varepsilon\right)$ through the following proposition.
Its proof is shown in Appendix \ref{sec:Proof-of-terProp}.
\begin{prop}
\label{prop:powerNepsi}There exists a minimum $\epsilon'>0$, such
that $\forall\sigma^{2}>\epsilon'$, the highest stable fixed point
of Eq. (\ref{eq:fixedpointEQ}) is $\Psi(\varepsilon/m\sum_{j\in[n]}\sigma_{j}^{2}+\sigma^{2})=\varepsilon/m\sum_{j\in[n]}\sigma_{j}^{2}+\sigma^{2}$.
\end{prop}
In the proposition, the highest fixed point is unique if $\partial\Psi(\widetilde{\tau}^{2})/\partial\widetilde{\tau}^{2}<1$
$\forall\widetilde{\tau}^{2}>0$, which means the increment of $\Psi(\widetilde{\tau}^{2})$
is never larger than that of $f(\widetilde{\tau}^{2})=\widetilde{\tau}^{2}$.
One implication of the proposition is, a stronger lattice reduction
method can help to make the fixed point smaller. E.g., with b-KZ,
one has 
\[
\sum_{j\in[n]}\sigma_{j}^{2}\leq\sum_{j\in[n]}\frac{\sqrt{j+3}}{2}\lambda_{j}(\mathbf{H})
\]
 for $n\geq2$. Another implication is, the performance of AMP should
be better if the real spark $\varepsilon$ is small. There is however
no genie granting which $\varepsilon$ fits the actual a priori knowledge.
According to our simulations, $\varepsilon=0.5$ is a good trade-off. 

For AMPG, similar analysis on fixed points can reveal the impact of
$\left\{ \sigma_{j}^{2}\right\} $ and prior variance $\sigma_{g}^{2}$.
By substituting Eq. (\ref{eq:-3}) to (\ref{eq:general thres}), the
fixed point function becomes

\begin{equation}
\Psi(\widetilde{\tau}^{2})=\sigma^{2}+\frac{1}{m}\sum_{j\in[n]}\frac{\widetilde{\tau}^{2}\sigma_{j}^{2}\sigma_{g}^{2}}{\widetilde{\tau}^{2}+\sigma_{j}^{2}\sigma_{g}^{2}}.\label{eq:gau_fixedpoint}
\end{equation}
 Let $\sigma_{\min}^{2}\triangleq\min_{j}\sigma_{j}^{2}$ and $\sigma_{\max}^{2}\triangleq\max_{j}\sigma_{j}^{2}$,
we have 
\[
\frac{n\widetilde{\tau}^{2}\sigma_{\min}^{2}\sigma_{g}^{2}}{m\left(\widetilde{\tau}^{2}+\sigma_{\min}^{2}\sigma_{g}^{2}\right)}\leq\Psi(\widetilde{\tau}^{2})-\sigma^{2}\leq\frac{n\widetilde{\tau}^{2}\sigma_{\max}^{2}\sigma_{g}^{2}}{m\left(\widetilde{\tau}^{2}+\sigma_{\max}^{2}\sigma_{g}^{2}\right)}.
\]
 As a consequence, one can easily prove that Eq. (\ref{eq:gau_fixedpoint})
has a unique stable fixed point that satisfies $\widetilde{\tau}^{2}\in[\widetilde{\tau}_{\min}^{2},\widetilde{\tau}_{\max}^{2}]$,
where 
\begin{align}
\widetilde{\tau}_{\min}^{2} & =\frac{1}{2}\left(\sigma^{2}+\left(\frac{n}{m}-1\right)\sigma_{\min}^{2}\sigma_{g}^{2}\right)\nonumber \\
 & +\frac{1}{2}\sqrt{\left(\sigma^{2}+\left(\frac{n}{m}-1\right)\sigma_{\min}^{2}\sigma_{g}^{2}\right)^{2}+4\sigma^{2}\sigma_{\min}^{2}\sigma_{g}^{2}},\label{eq:taufixed}
\end{align}
 and $\widetilde{\tau}_{\max}^{2}$ is defined by replacing $\sigma_{\min}^{2}$
with $\sigma_{\max}^{2}$ in (\ref{eq:taufixed}). In order to make
the fixed point small, one should also make the lattice basis short.
The setting of $\sigma_{g}^{2}$ is also a trade-off: it should be
set smaller to yield a lower fixed point, but there should be a minimum
for it so that the imposed signal distribution still reflects discrete
Gaussian information. A general principle for finding the trade-off
value is left as an open question.

\subsection{\label{sub:Complexity}Complexity of AMP}

The complexity is assessed by counting the number of floating-point
operations (flops). For the threshold functions (\ref{eq:SPAeta})
(\ref{eq:spakapa}) of AMPT, we can use $\sinh\left(\frac{u}{v}\right)\approx\frac{u}{v},$
$\cosh\left(\frac{u}{v}\right)\approx1+\frac{u^{2}}{2v^{2}}$ for
small $u/v$ since $\sinh x=\sum_{k=0}^{\infty}\frac{x^{2k}}{\left(2k\right)!}$,
$\cosh x=\sum_{k=0}^{\infty}\frac{x^{2k+1}}{\left(2k+1\right)!}$.
Outer bounding (\ref{eq:SPAeta}) (\ref{eq:spakapa}) is also possible
for large $u/v$, so we can approximate (\ref{eq:SPAeta}) (\ref{eq:spakapa})
by $O(1)$ flops. The $O(1)$ complexity also holds for (\ref{eq:-3})
(\ref{eq:-4}) of AMPG. In conclusion, the complexity of our AMP algorithm
is $O(mnT)$. On the contrary, a full enumeration with a ternary constraint
already requires at least $O(3^{n})$ flops, and ZF/SIC requires $O(mn^{2})$
flops.

\section{\label{secamp:Simulations}Simulations}

In this section, the symbol error rate (SER) and complexity performance
of the proposed hybrid precoding scheme are examined through Monte
Carlo simulations. The impacts of chosen parameters in the threshold
functions are also studied. For comparison, the sphere precoding method
\cite{Peel2005,Hochwald2005} and LR aided precoding methods based
on ZF/SIC are also tested. Throughout this section, b-LLL with list
size $1$ is adopted as the default LR option, and we refer to \cite{Lyu2017}
for a full comparison of different reduction algorithms. In all the
AMP algorithms, we set $\sigma^{2}=\left\Vert \mathbf{y}-\mathbf{H}\hat{\mathbf{x}}\right\Vert ^{2}/m^{1.5}$
(so as to approximate $\left\Vert \mathbf{y}-\mathbf{H}\mathbf{x}^{\mathrm{cvp}}\right\Vert ^{2}/m$),
and $T=20$.

\begin{figure}[t]
\center

\includegraphics[clip,width=3.4in]{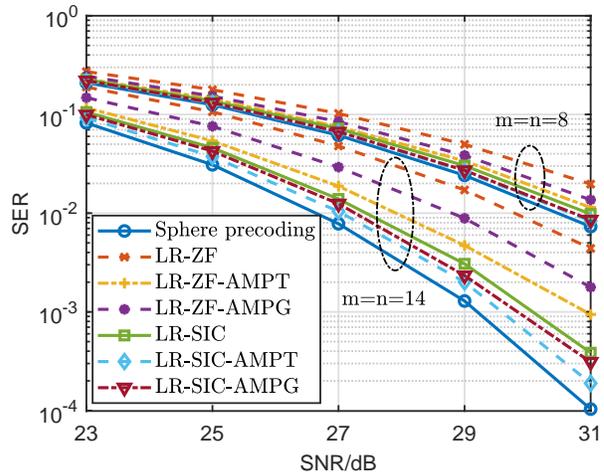}

\protect\caption{The symbol error rate of different algorithms.}
\label{figSER8n14} 
\end{figure}

First, Fig. \ref{figSER8n14} illustrates the SNR versus SER performance
of different algorithms using a modulation size $M=32$ for antenna
configurations $m=n=8$ and $m=n=14$. We set $\varepsilon=0.5$ in
AMPT and $\sigma_{g}^{2}=2$ in AMPG. As shown in the figure, both
AMPT and AMPG can improve the performance of LR-ZF/SIC towards that
of sphere precoding. These gains become more evident as the size of
the system grows from $m=n=8$ to $m=n=14$, in which AMPT improves
LR-ZF by $3$dB and LR-SIC by $0.8$dB. 

\begin{figure}[t]
\center

\includegraphics[clip,width=3.4in]{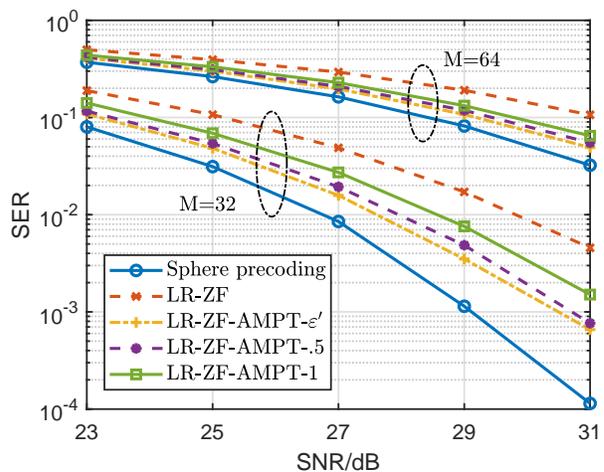}

\protect\caption{The impact of spark values in AMPT with $m=n=14$.}
\label{figSpark} 
\end{figure}

Next, we examine the effect of choosing different spark values in
AMPT, with $M=32,64$ and $m=n=14$. The AMPT algorithm using the
real spark (by comparing to sphere precoding) to noted as AMPT-$\varepsilon'$.
Two other references are $\varepsilon=0.5$ and $\varepsilon=1$.
According to Fig. \ref{figSpark}, the idealised AMPT-$\varepsilon'$
performs $1$dB better than AMPT-$1$, but is within $0.2$dB distance
to AMPT-$.5$. This suggests that in practice we can adopt $\varepsilon=0.5$
as a reasonable configuration.

\begin{figure}[t]
\center

\includegraphics[clip,width=3.4in]{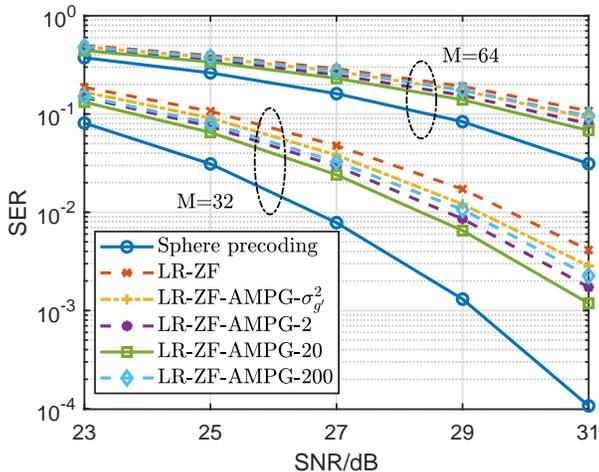}

\protect\caption{The impact of variance $\sigma_{g}^{2}$ in AMPG with $m=n=14$.}
\label{figAMPvar} 
\end{figure}

Similarly, the effect of chosen variance $\sigma_{g}^{2}$ in AMPG
is studied in Fig. \ref{figAMPvar}. The suffixes after AMPG refer
to setting $\sigma_{g}^{2}$ as $\sigma_{g'}^{2}=\left\Vert \mathbf{x}^{\mathrm{cvp}}-\hat{\mathbf{x}}\right\Vert ^{2}/n$,
and $2$, $20$, $200$, respectively. Other configurations are identical
to those in Fig. \ref{figSpark}. An observation from Fig. \ref{figAMPvar}
is that the AMPG-$\sigma_{g'}^{2}$ algorithm is not better than those
with manually chosen variances; this refeclts the fact that $\sigma_{g}^{2}$
can not be too small so as to reflect discrete Gaussian information
(c.f. Section \ref{sub:The-impact-of_power}). In addition, the trade-offs
$\sigma_{g}^{2}=2,20$ work better than the too large value $\sigma_{g}^{2}=200$
and the too small value ($\sigma_{g}^{2}=\sigma_{g'}^{2}$). 

\begin{figure}[t]
\center

\includegraphics[clip,width=3.4in]{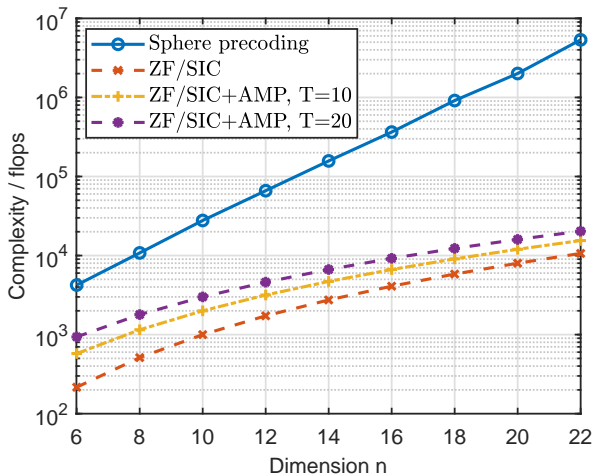}

\protect\caption{The complexity of different algorithms.}
\label{figComplexity} 
\end{figure}

In the last example, we examine the complexity of our AMP algorithms.
We use estimations in Section \ref{sub:Complexity} to measure the
complexity of ZF/SIC and AMP. As for the sphere decoding algorithm,
it is implemented after b-LLL so as to decrease its complexity. All
algorithms can take the benefits of b-LLL, and the complexity costed
by lattice reduction is not counted for all of them. The actual complexity
of sphere precoding depends on the inputs, so we count the number
of nodes it visited, and assign $2k+7$ flops to a visited node in
layer $k$ \cite{Lyu2017}. From Fig. \ref{figComplexity}, we can
see that AMP with a constant iteration number, e.g., $T=10$ or $T=20$,
is adding little complexity budget to that of ZF/SIC. On the contrary,
the exponential complexity of sphere decoding makes it at least $200$
times more complicated than our ZF/SIC+AMP scheme in dimension $n=22$. 

\section{\label{sec:Extension-to-Data}Extension to Data Detection in Massive
MIMO}

The developed hybrid precoding (decoding) scheme for VP can be directly
extended to address the data detection problem in small-scaled MIMO
systems whose underlying CVP has more constraints. We omit the presentation
of similar results. The more interesting extension that we will pursuit
in this section is to data detection in massive MIMO systems, where
the base stations are equipped with hundreds of antennas to simultaneously
serve tens of users \cite{Larsson2014}. In the classical i.i.d. frequency-flat
Rayleigh fading MIMO channels, the set-up of massive MIMO implies
that channel matrix is extremely tall in the corresponding CVP. As
a result, we can regard these lattice bases (channel matrices) that
are short and orthogonal as naturally reduced. \textit{This suggests
we can apply our hybrid scheme to massive MIMO without using lattice
reduction.}

\subsection{System Model and the Reduced Basis}

With a slight abuse of notation, we write the system model in the
uplink of massive MIMO as 
\begin{equation}
\mathbf{y}=\mathbf{H}\mathbf{x}+\mathbf{w},\label{eq:detectionmodel}
\end{equation}
where $\mathbf{y}\in\mathbb{R}^{m}$ is the received signal vector
at the base station, $\mathbf{H}\in\mathbb{R}^{m\times n}$ denotes
the channel matrix whose entries follow the distribution of $\mathrm{N}(0,1)$,
$\mathbf{w}\in\mathbb{R}^{m}$ is the additive noise vector whose
entries admit $\mathrm{N}(0,\sigma^{2})$, and $\mathbf{x}\in\mathcal{M}^{n}$
is the transmitted signal vector that contains the data symbols from
all the user terminals. For ease of presentation, we set the constellation
as $\mathcal{M}=\{-M,-M+1,\ldots,M-1,M\}$. The special constraint
in massive MIMO is that $m\gg n$, based on which the simple LMMSE
detection suffices to provide near-optimal performance. Let $\sigma_{s}^{2}$
denote the averaged power of $\mathbf{x}$, by using LMMSE equalization
we have
\[
\mathbf{x}^{\mathrm{lmmse}}=\lfloor\left(\mathbf{H}^{\top}\mathbf{H}+\sigma^{2}/\sigma_{s}^{2}\mathbf{I}_{n}\right)^{-1}\mathbf{H}^{\top}\mathbf{y}\rceil.
\]
 It is well known that LMMSE is a variant of ZF and they become equivalent
as $\sigma^{2}\rightarrow0$, and its computational complexity is
$O\left(n^{3}+mn^{2}\right)$.

\begin{table*}[tbh]
\protect\caption{The values $\hat{x}_{i}-x_{i}^{\mathrm{cvp}}$ with $i\in[n]$ and
their probabilities in data detection.}
\label{tab:detection_error} \centering %
\begin{tabular}{|c|c||c|c|c|c|c|c|c|c|c|}
\hline 
\multicolumn{1}{|c}{error} & distance & $-4$ & $-3$ & $-2$ & $-1$ & $0$ & $1$ & $2$ & $3$ & $4$\tabularnewline
\hline 
$m=16,n=8$ & ZF & $0$ & $0$ & $0.0003$ & $0.0796$ & $0.8458$ & $0.0744$ & $0$ & $0$ & $0$\tabularnewline
\cline{2-11} 
$\mathrm{SNR}=10\mathrm{dB}$ & LR-ZF & $0$ & $0$ & $0.0021$ & $0.0825$ & $0.8285$ & $0.0854$ & $0.0015$ & $0$ & $0$\tabularnewline
\hline 
$m=8,n=8$ & ZF & $0.0090$ & $0.0174$ & $0.0401$ & $0.1594$ & $0.5035$ & $0.1544$ & $0.0409$ & $0.0163$ & $0.0077$\tabularnewline
\cline{2-11} 
$\mathrm{SNR}=10\mathrm{dB}$ & LR-ZF & $0.0086$ & $0.0166$ & $0.0369$ & $0.1090$ & $0.6105$ & $0.1087$ & $0.0334$ & $0.0160$ & $0.0103$\tabularnewline
\hline 
$m=16,n=8$ & ZF & $0$ & $0$ & $0$ & $0.0003$ & $0.9995$ & $0.0002$ & $0$ & $0$ & $0$\tabularnewline
\cline{2-11} 
$\mathrm{SNR}=30\mathrm{dB}$ & LR-ZF & $0$ & $0$ & $0$ & $0.0001$ & $0.9999$ & $0$ & $0$ & $0$ & $0$\tabularnewline
\hline 
$m=8,n=8$ & ZF & $0.0057$ & $0.0077$ & $0.0195$ & $0.1074$ & $0.6937$ & $0.1046$ & $0.0176$ & $0.0079$ & $0.0043$\tabularnewline
\cline{2-11} 
$\mathrm{SNR}=30\mathrm{dB}$ & LR-ZF & $0.0018$ & $0.0027$ & $0.0037$ & $0.0181$ & $0.9299$ & $0.0170$ & $0.0056$ & $0.0032$ & $0.0022$\tabularnewline
\hline 
\end{tabular}
\end{table*}

Here, we notice that channel matrices in massive MIMO represent extremely
good lattice bases. For instance, a tall channel matrix with dimension
$2n\times n$ already represents a lattice basis that often outcompetes
boosted KZ (to our knowledge, this is almost the strongest lattice
reduction). To support this argument, we show the symbol-wise error
distance in decoding (\ref{eq:detectionmodel}) by using LR (boosted
KZ) aided ZF and ZF. Table \ref{tab:detection_error} reveals this
result using $M=14$, $\left(m,n\right)=\left(16,8\right)$, $\left(m,n\right)=(8,8)$,
$\mathrm{SNR}=10\mathrm{dB}$ and $\mathrm{SNR}=30\mathrm{dB}$. We
have the following observations from the table: For a square channel
matrix with $m=n$, LR-ZF indeed has smaller error ratios than those
of ZF. But as the channel matrix becomes tall with $m=2n$, ZF performs
close to LR-ZF ($\mathrm{SNR}=30\mathrm{dB}$) or even outperforms
LR-ZF ($\mathrm{SNR}=10\mathrm{dB}$). Similar observations can also
be made for other sizes of constellations, SNRs and sizes of the system.

This phenomenon is however not a surprise: as $m/n$ grows larger,
the vectors $\mathbf{h}_{1},\ldots,\mathbf{h}_{n}$ in the basis become
almost mutually orthogonal. Since any linear combination of these
vectors can only be longer, $\mathbf{h}_{1},\ldots,\mathbf{h}_{n}$
would become the shortest $n$ independent vectors of $\mathcal{L}(\mathbf{H})$
and we have $\left\Vert \mathbf{h}_{i}\right\Vert =\lambda_{i}(\mathbf{H})$
for $i\in[n]$. Compared to boosted KZ which only upper bounds $\left\Vert \mathbf{h}_{i}\right\Vert $
to $O(\sqrt{i})\lambda_{i}(\mathbf{H})$, these shortest independent
vectors are much more desirable.

\subsection{Simulations}

To see the advantage of using hybrid decoding in massive MIMO, we
run simulations to obtain SERs for different algorithms. With a relatively
large constellation size, the AMP algorithm using exact a priori knowledge
no longer suits our problem as it is slow, unstable and divergent
\cite{Jeon2015,Jeon2016}. It is therefore reasonable to adopt AMPG
\cite{Bayati2011,lyu15,Chen2017} as a benchmark, because it has the
best convergence behavior among all AMP-based algorithms without using
hybrid decoding. Another benchmark is the LMMSE estimator, and we
will use AMPG or AMPT (both have complexity $O(mnT)$) as the algorithm
in ``Phase 2'' based on it. 

\begin{figure}[t]
\center

\includegraphics[clip,width=3.4in]{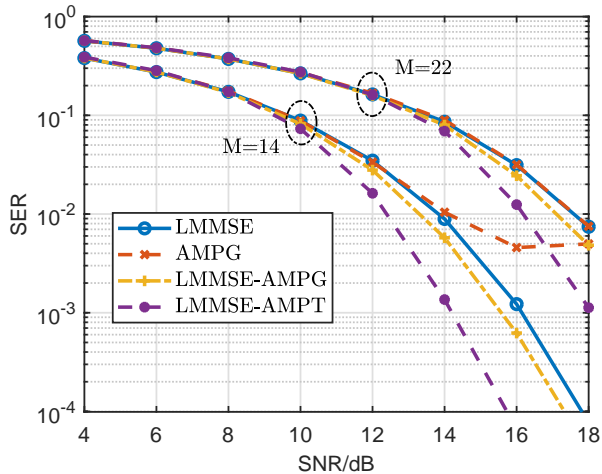}

\protect\caption{The symbol error rate of different algorithms with $m=128$, $n=48$.}
\label{figdetecSER48} 
\end{figure}

\begin{figure}[t]
\center

\includegraphics[clip,width=3.4in]{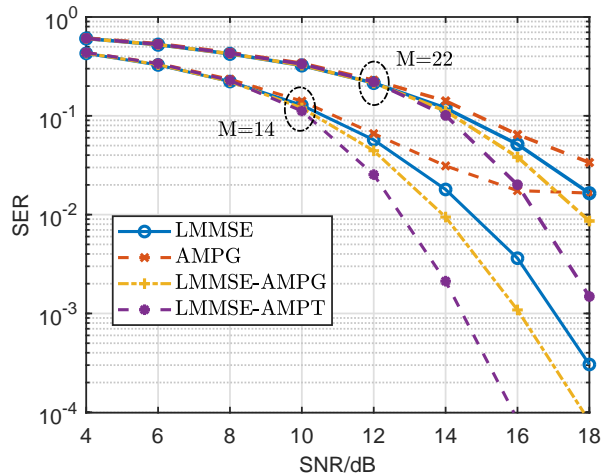}

\protect\caption{The symbol error rate of different algorithms with $m=128$, $n=64$.}
\label{figdetecSER64} 
\end{figure}

The SERs of these algorithms are shown in Figs. \ref{figdetecSER48}
and \ref{figdetecSER64}, with constellation size $M=14,22$ and system
dimension $(m,n)=(128,48),(128,64)$. As revealed in the figures,
both AMPG and AMPT can improve the performance of LMMSE to a certain
degree, but the improvement of AMPT is more evident as its threshold
functions are non-linear. Note that the hybrid scheme can also employ
AMPG as the first-round algorithm to make the total complexity of
hybrid decoding as $O(mnT)$, but the bad performance of AMPG at high
SNR dictates the overall performance.

\section{Conclusions}

In this work, we have presented a hybrid precoding scheme for VP.
The precoding problem in VP is about solving CVP in a lattice, and
this problem is quite general because the signal space lies in integers
$\mathbb{Z}$. After performing LR aided ZF/SIC, we indicated that
the signal space had been significantly reduced, and this information
paved the way for the application of the celebrated AMP algorithm.
Considering ternary distributions and Gaussian distributions, we have
designed threshold functions that have closed-form expressions. Our
simulations showed that attaching AMP to LR-ZF or LR-SIC can provide
around $0.5$dB to $2.5$dB gain in SER for VP, where the AMP algorithm
only incurred complexity in the order of $O(mnT)$. Lastly, we have
also demonstrated that the hybrid scheme can be extended to data detection
in massive MIMO without using lattice reduction.

\appendices{}

\section{\label{sec:ODmu}On using reduced bases for AMP}

In our precoding problem, the mixing matrix $\mathbf{H}$ comes from
lattice reduction rather than naturally having i.i.d. Gaussian entries.
Although $\mathbf{H}$ is known to be short and nearly orthogonal
after lattice reduction, its statistical information cannot be exactly
analyzed by only using the theory of lattices. To provide some complements
to our simulation results that have confirmed the feasibility of using
reduced bases for AMP, our aim in this section is to explain why the
reduce bases can work in principle.

\begin{figure}[th]
\center

\includegraphics[clip,width=3.4in]{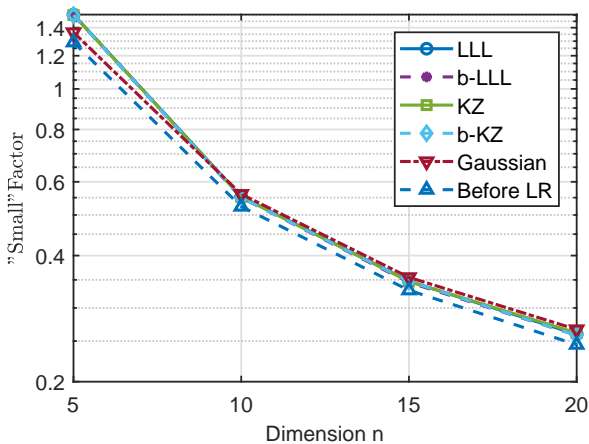}

\protect\caption{The averaged ``small'' factors of different algorithms for $\mathbf{H}\in\mathbb{R}^{n\times n}$.}
\label{figsmall} 
\end{figure}

The first reason is that all the edges on the bipartite graph are
weak for a reduced basis. It was suggested by Rangan et al. \cite{Rangan2017}
that the AMP-style approximations are effective if the messages are
propagating on weak edges. In their definition \cite[P. 4578]{Rangan2017},
the entries of a mixing matrix $\mathbf{H}$ are called ``small''
if no individual component can have a significant effect on the row-sum
or column-sum of $\mathbf{H}$. Here we define a ``small'' factor
to measure this effect:
\[
\mu_{s}\left(\mathbf{H}\right)\triangleq\max_{i\in[m],j\in[n]}\left(\max\left(\frac{|H_{i,j}|}{\sum_{i}|H_{i,j}|},\frac{|H_{i,j}|}{\sum_{j}|H_{i,j}|}\right)\right).
\]
 In Fig. \ref{figsmall}, we plot the averaged ``small'' factors
$\mathbb{E}\mu_{s}\left(\mathbf{H}\right)$ produced by different
methods. The lattice reduction methods, noted as ``LLL'', ``b-LLL'',
``KZ'' and ``b-KZ'' , are applied on the inverse of Gaussian random
matrices of rank $n$. The ``small'' factors of Gaussian random
matrices with $\mathrm{N}(0,1)$ entries, noted as ``Gaussian'',
and those before lattice reduction, noted as ``Before LR'', are
also included for comparison. One thing we can observe from the figure
is that the lattice reduction methods behave as good as Gaussian entries.
We also note the figure shows the bases before lattice reduction already
exhibit rather weak edges, but this does not suggest they correspond
to more efficient AMP methods. We notice that a small $\mu_{s}\left(\mathbf{H}\right)$
is only a necessary condition for AMP. Specifically, for a matrix
with $H_{i,j}=1,\thinspace\forall i,j$, we have $\mu_{s}\left(\mathbf{H}\right)=1/n$
that is arbitrarily small while $\mathbf{H}$ is ill-conditioned. 

\begin{figure}[th]
\center

\includegraphics[clip,width=3.4in]{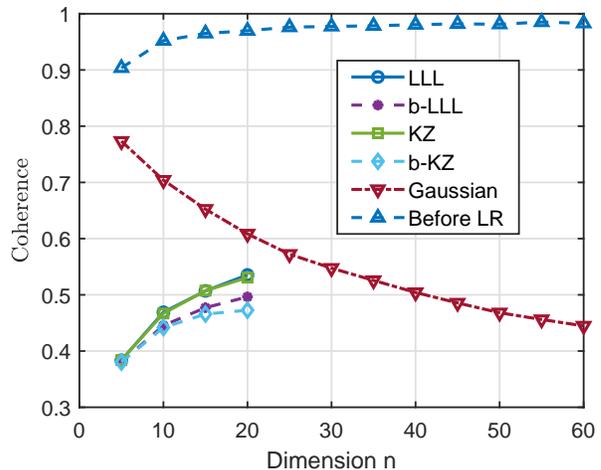}

\protect\caption{The coherence parameters of different algorithms for $\mathbf{H}\in\mathbb{R}^{n\times n}$.}
\label{figcoh} 
\end{figure}

The second reason is that a reduced basis has a small coherence parameter
\cite{Foucart2013b} defined by 
\[
\mu_{c}\left(\mathbf{H}\right)\triangleq\max_{1\leq i\neq j\leq n}|\mathbf{h}_{i}^{\top}\mathbf{h}_{j}|/\left\Vert \mathbf{h}_{i}\right\Vert \left\Vert \mathbf{h}_{j}\right\Vert ,
\]
where $\mathbf{H}=\left[\mathbf{h}_{1},\ldots\thinspace,\mathbf{h}_{n}\right]$.
This metric can reflect the column-wise independence. In Fig. \ref{figcoh},
we plot the expected coherence parameter $\mathbb{E}\mu_{c}\left(\mathbf{H}\right)$
of different lattice reduction algorithms from dimensions $5$ to
$20$, and include Gaussian bases and the bases before LR for comparison.
Other settings are the same as those in Fig. \ref{figsmall}. As shown
in Fig. \ref{figcoh}, the coherence parameters can be significantly
reduced after using lattice reduction. Most importantly, Fig. \ref{figcoh}
suggests that a coherence parameter of $\mu_{c}\left(\mathbf{H}\right)=0.5$
that corresponds to a Gaussian random matrix of dimension $n=40$
is equivalent to those of lattice reduction with much smaller dimensions,
e.g., $n=15$ with LLL and $n=20$ with boosted LLL.

\section{\label{sec: short quan}Proof of Eq. (\ref{eq:kzsic}) in Theorem
\ref{thm:thmlll}}

When proving the energy efficiency of b-KZ aided SIC/ZF, the following
lemma would be needed. Remind that $\mathbf{H}=\mathbf{QR}$ is the
QR factorization. 

\begin{lem}[\cite{Lyu2017}]
\label{prop:gs lambda1 }Suppose a basis $\mathbf{H}$ is b-KZ reduced,
then this basis conforms to
\begin{equation}
\lambda_{1}(\mathbf{H})^{2}\leq\frac{8i}{9}(i-1)^{\ln(i-1)/2}R_{i,i}^{2},\label{eq:lambda1 gslen}
\end{equation}
\begin{equation}
\left\Vert \mathbf{h}_{i}\right\Vert ^{2}\leq\Big(1+\frac{2i}{9}(i-1)^{1+\ln(i-1)/2}\Big)R_{i,i}^{2}\label{eq:len gs3}
\end{equation}
for $1\leq i\leq n$, and 
\begin{equation}
R_{k-j+1,k-j+1}^{2}\leq\frac{8j}{9}(j-1)^{\ln(j-1)/2}R_{k,k}^{2}\label{eq:r_ratio}
\end{equation}
 for $2\leq k\leq n$, $j\leq k$.
\end{lem}
Under the unitary transform $\mathbf{Q}^{\top}$, we aim to prove
an equivalence of (\ref{eq:etaCVP}) as 
\begin{equation}
\left\Vert \mathbf{\bar{y}}-\mathbf{R}\hat{\mathbf{x}}\right\Vert \leq\eta_{n}\min_{\mathbf{x}\in\mathbb{Z}^{n}}\left\Vert \mathbf{\bar{\mathbf{y}}}-\mathbf{Rx}\right\Vert ,\label{eq:basiceq}
\end{equation}
with $\mathbf{\mathbf{\bar{y}}}=\mathbf{Q}^{\top}\mathbf{y}$. Let
$\mathbf{v}^{\mathrm{cvp}}=\mathbf{R}\mathbf{x}^{\mathrm{cvp}}$ be
the closest vector to $\mathbf{\mathbf{\bar{y}}}$, and $\mathbf{v}^{\mathrm{sic}}=\mathbf{R}\mathbf{x}^{\mathrm{sic}}$
be the vector founded by SIC. As the SIC parallelepiped generally
mismatches the Voronoi region, we need to investigate the relation
of $x_{n}^{\mathrm{cvp}}$ and $x_{n}^{\mathrm{sic}}=\lfloor\bar{y}_{n}/R_{n,n}\rceil$
as in that in \cite{Babai1986}. If $x_{n}^{\mathrm{cvp}}=x_{n}^{\mathrm{sic}}$,
we only need to investigate $\eta_{n-1}$ in another $n-1$ dimensional
CVP by setting $\mathbf{\bar{y}}\leftarrow\mathbf{\bar{y}}-\mathbf{r}_{n}x_{n}^{\mathrm{sic}}$:
$\left\Vert \mathbf{\bar{y}}-\mathbf{R}_{1:n,1:n-1}\mathbf{x}_{1:n-1}^{\mathrm{sic}}\right\Vert \leq\eta_{n-1}\min_{\mathbf{x}\in\mathbb{Z}^{n-1}}\left\Vert \mathbf{\bar{y}}-\mathbf{R}_{1:n,1:n-1}\mathbf{x}\right\Vert $.
When this situation continues till the first layer, one clearly has
$\eta_{1}=1$. Generally, we can assume that this mismatch first happens
in the $k$th layer, i.e., assume $x_{k}^{\mathrm{cvp}}\neq x_{k}^{\mathrm{sic}},$
$k\in\{2,\ldots\thinspace,n\}$, then $|\bar{y}_{k}/R_{k,k}-x_{k}^{\mathrm{cvp}}|\geq\frac{1}{2}$,
and 
\begin{equation}
\left\Vert \bar{\mathbf{y}}-\mathbf{v}^{\mathrm{cvp}}\right\Vert ^{2}\geq r_{k,k}^{2}(\bar{y}_{k}/R_{k,k}-x_{k}^{\mathrm{cvp}})^{2}\geq R_{k,k}^{2}/4.\label{eq:func2}
\end{equation}
According to (\ref{eq:r_ratio}) of b-KZ, we have $R_{k-j+1,k-j+1}^{2}\leq\frac{8j}{9}(j-1)^{\ln(j-1)/2}R_{k,k}^{2}$,
then the SIC solution $\mathbf{R}_{1:n,1:k}\mathbf{x}_{1:k}^{\mathrm{sic}}$
satisfies 
\begin{align}
\left\Vert \bar{\mathbf{y}}-\mathbf{R}_{1:n,1:k}\mathbf{x}_{1:k}^{\mathrm{sic}}\right\Vert ^{2} & \le\frac{1}{4}\sum_{i=1}^{k}R_{i,i}^{2}\nonumber \\
 & \leq\left(\frac{1}{4}+\frac{2k}{9}\left(k-1\right)^{1+\ln(k-1)/2}\right)R_{k,k}^{2}.\label{eq:func1}
\end{align}
Combining (\ref{eq:func1}) and (\ref{eq:func2}), and choose $k=n$
in the worst case, we have
\[
\left\Vert \bar{\mathbf{y}}-\mathbf{v}^{\mathrm{sic}}\right\Vert ^{2}\leq\left(1+\frac{8n}{9}\left(n-1\right)^{1+\ln(n-1)/2}\right)\left\Vert \bar{\mathbf{y}}-\mathbf{v}^{\mathrm{cvp}}\right\Vert ^{2}.
\]

\section{\label{sec:Proof-of-zflr}Proof of Theorem \ref{thm:thmZFlr}}

The energy efficiency of b-LLL/b-KZ aided ZF precoding is non-trivial
to prove because we cannot employ the size reduction conditions to
claim an upper bound for $(\mathbf{A}^{i})_{1,1}^{-1}$ as that in
\cite[Eq. (65)]{Ling2011}, in which $\mathbf{A}^{i}=\mathbf{R}_{i:n,i:n}^{\top}\mathbf{R}_{i:n,i:n}$.
This condition is crucial as one already has 
\[
\sin^{2}\theta_{i}=\frac{1}{\left\Vert \mathbf{h}_{i}\right\Vert ^{2}(\mathbf{A}^{i})_{1,1}^{-1}}
\]
according to \cite[Appx. I]{Ling2011}, where $\theta_{i}$ is the
angle between $\mathbf{h}_{i}$ and $\mathrm{span}(\mathbf{h}_{1},\ldots,\thinspace\mathbf{h}_{i-1},\mathbf{h}_{i+1},\ldots,\thinspace\mathbf{h}_{n})$.
The following lemma proves a lower bound for $\sin^{2}\theta_{i}$
by only invoking the relation between $\left\Vert \mathbf{h}_{i}\right\Vert ^{2}$
and $R_{i,i}^{2}$. 
\begin{lem}
\label{lem:theta}Let $\mathbf{H}$ be a b-KZ reduced basis, then
it satisfies $\sin^{2}\theta_{i}\geq\left(\prod_{k=i}^{n}k^{2+\ln(k)/2}\right)^{-1}.$\end{lem}
\begin{IEEEproof}
Define $\mathbf{M}^{k}=\mathbf{R}_{i:k,i:k}^{-1}$ along with $\mathbf{M}^{i}=R_{i,i}^{-1}$,
then 
\[
\mathbf{M}^{k}=\left[\begin{array}{cc}
\mathbf{M}^{k-1} & R_{k,k}^{-1}\mathbf{M}^{k-1}\mathbf{R}_{i:k-1,k}\\
\mathbf{0} & R_{k,k}^{-1}
\end{array}\right].
\]
 By using Cauchy\textendash Schwarz inequality on $\mathbf{M}_{1,:}^{k-1}\mathbf{R}_{i:k-1,k}$,
we also have 
\begin{align}
\left\Vert \mathbf{M}_{1,:}^{k}\right\Vert ^{2} & =\left\Vert \mathbf{M}_{1,:}^{k-1}\right\Vert ^{2}+\left(R_{k,k}^{-1}\mathbf{M}_{1,:}^{k-1}\mathbf{R}_{i:k-1,k}\right)^{2}\nonumber \\
 & \leq\left\Vert \mathbf{M}_{1,:}^{k-1}\right\Vert ^{2}\left(1+R_{k,k}^{-2}\left\Vert \mathbf{R}_{i:k-1,k}\right\Vert ^{2}\right).\label{eq:bkz prom}
\end{align}
It is evident that $\left\Vert \mathbf{R}_{i:k-1,k}\right\Vert ^{2}\leq\left\Vert \mathbf{h}_{k}\right\Vert ^{2}-R_{k,k}^{2}\overset{(a)}{\leq}\Big(1+\frac{2k}{9}(k-1)^{1+\ln(k-1)/2}\Big)R_{k,k}^{2}-R_{k,k}^{2}$,
where (a) is due to inequality (\ref{eq:len gs3}), so that $R_{k,k}^{-2}\left\Vert \mathbf{R}_{i:k-1,k}\right\Vert ^{2}\leq\frac{2k}{9}(k-1)^{1+\ln(k-1)/2}$.
Substitute this into (\ref{eq:bkz prom}), then
\begin{align*}
\left\Vert \mathbf{M}_{1,:}^{k}\right\Vert ^{2} & \leq\left\Vert \mathbf{M}_{1,:}^{k-1}\right\Vert ^{2}\left(1+\frac{2k}{9}(k-1)^{1+\ln(k-1)/2}\right)\\
 & \leq\left\Vert \mathbf{M}_{1,:}^{k-1}\right\Vert ^{2}k^{2+\ln(k)/2}.
\end{align*}
 By induction, one has
\[
(\mathbf{A}^{i})_{1,1}^{-1}=\left\Vert \mathbf{M}_{1,:}^{n}\right\Vert ^{2}\leq R_{i,i}^{-2}\prod_{k=i+1}^{n}k^{2+\ln(k)/2}.
\]
and thus 
\[
\sin^{2}\theta_{i}\geq\frac{R_{i,i}^{2}}{\left\Vert \mathbf{h}_{i}\right\Vert ^{2}\prod_{k=i+1}^{n}k^{2+\ln(k)/2}}\geq\left(\prod_{k=i}^{n}k^{2+\ln(k)/2}\right)^{-1},
\]
where the second inequality is due to Lem. \ref{prop:gs lambda1 }.
\end{IEEEproof}
With the same technique as above, we can bound $\sin^{2}\theta_{i}$
for b-LLL.
\begin{lem}
\label{lem:theta-1}Let $\mathbf{H}$ be a b-LLL reduced basis, then
it satisfies $\sin^{2}\theta_{i}\geq\left(\prod_{k=i}^{n}\beta^{k-1}\right)^{-1}$.
\end{lem}
We proceed to investigate inequality (\ref{eq:basiceq}). Let $\mathbf{v}^{\mathrm{cvp}}=\mathbf{R}\mathbf{x}^{\mathrm{cvp}}$
be the closest vector to $\mathbf{\mathbf{\bar{y}}}$, and $\mathbf{v}^{\mathrm{zf}}=\mathbf{R}\mathbf{x}^{\mathrm{zf}}$
be the vector found by ZF. Define $\mathbf{v}^{\mathrm{cvp}}-\mathbf{v}^{\mathrm{zf}}=\sum_{i=1}^{n}\phi_{i}\mathbf{h}_{i}$
with $\phi_{i}\in\mathbb{Z}$. If $\mathbf{v}^{\mathrm{cvp}}=\mathbf{v}^{\mathrm{zf}}$,
then the energy efficiency $\eta_{n}=1$. If $\mathbf{v}^{\mathrm{cvp}}\neq\mathbf{v}^{\mathrm{zf}},$
then 
\[
\left\Vert \mathbf{v}^{\mathrm{cvp}}-\mathbf{v}^{\mathrm{zf}}\right\Vert \leq\sum_{j=1}^{n}\left\Vert \phi_{j}\mathbf{h}_{j}\right\Vert .
\]
At the same time, we have 
\begin{align*}
\mathbf{v}^{\mathrm{cvp}}-\bar{\mathbf{y}} & =\mathbf{v}^{\mathrm{cvp}}-\mathbf{v}^{\mathrm{zf}}+\mathbf{v}^{\mathrm{zf}}-\bar{\mathbf{y}}\\
 & =(\phi_{k}+\phi_{k}^{\mathrm{zf}})\mathbf{h}_{k}+\mathbf{m}',
\end{align*}
where $\mathbf{m}'\in\mathrm{span}(\mathbf{h}_{1},\ldots,\thinspace\mathbf{h}_{k-1},\mathbf{h}_{k+1},\ldots,\thinspace\mathbf{h}_{n})$,
$\mathbf{v}^{\mathrm{zf}}-\bar{\mathbf{y}}=\sum_{i=1}^{n}\phi_{i}^{\mathrm{zf}}\mathbf{h}_{i}$
satisfies $|\phi_{i}^{\mathrm{zf}}|\leq1/2$ $\forall i$, and $k\triangleq\arg\max_{i}\left\Vert \phi_{i}\mathbf{h}_{i}\right\Vert $.
From Lem. \ref{lem:theta}, $\left\Vert (\phi_{k}+\phi_{k}^{\mathrm{zf}})\mathbf{h}_{k}+\mathbf{m}'\right\Vert \geq|\phi_{k}+\phi_{k}^{\mathrm{zf}}|\left(\prod_{j=k}^{n}j^{2+\ln(j)/2}\right)^{-1}\left\Vert \mathbf{h}_{k}\right\Vert $,
so that 
\[
\left\Vert \mathbf{v}^{\mathrm{cvp}}-\bar{\mathbf{y}}\right\Vert \geq|\phi_{k}|\left(2\prod_{j=k}^{n}j^{2+\ln(j)/2}\right)^{-1}\left\Vert \mathbf{h}_{k}\right\Vert 
\]
 as $|\phi_{k}+\phi_{k}^{\mathrm{zf}}|\geq|\phi_{k}|/2$. According
to the triangle inequality, one has for b-KZ that
\begin{align*}
\left\Vert \mathbf{v}^{\mathrm{zf}}-\bar{\mathbf{y}}\right\Vert  & \leq\left\Vert \mathbf{v}^{\mathrm{zf}}-\mathbf{v}^{\mathrm{cvp}}\right\Vert +\left\Vert \mathbf{v}^{\mathrm{cvp}}-\bar{\mathbf{y}}\right\Vert \\
 & \leq\left(2n\prod_{j=1}^{n}j^{2+\ln(j)/2}+1\right)\left\Vert \mathbf{v}^{\mathrm{cvp}}-\bar{\mathbf{y}}\right\Vert .
\end{align*}
 One can similarly prove for b-LLL that 
\[
\left\Vert \mathbf{v}^{\mathrm{zf}}-\bar{\mathbf{y}}\right\Vert \leq\left(2n\prod_{j=1}^{n}\beta^{j-1}+1\right)\left\Vert \mathbf{v}^{\mathrm{cvp}}-\bar{\mathbf{y}}\right\Vert .
\]

\section{\label{sec:Proof-of-se}Proof of Theorem \ref{thm:SE non-uniform power}}
\begin{IEEEproof}
We follow \cite[Sec. 1.3]{Bayati2011} to analysis the state evolution
equation (\ref{eq:se standard}). Let the observation equation be
$\mathbf{y}^{t}=\mathbf{H}^{t}\bar{\mathbf{x}}+\mathbf{w}$, where
the distribution of $\bar{\mathbf{x}}$ is denoted by $p_{X}$, $H_{b,i}\sim\mathrm{N}(0,\sigma_{i}^{2}/m)$,
and $w_{i}\in\mathrm{N}(0,\sigma^{2})$. Without the Onsager term,
the residual equation becomes:

\begin{equation}
\mathbf{r}^{t}=\mathbf{y}^{t}-\mathbf{H}^{t}\mathbf{x}^{t}.\label{eq:26-1-1-1-1-1}
\end{equation}
Along with with independently generated $\left\{ \mathbf{H}^{t}\right\} $,
the estimation equation becomes: 

\begin{equation}
\mathbf{x}^{t+1}=\eta(\Theta\mathbf{H}^{t\top}\mathbf{r}^{t}+\mathbf{x}^{t},\Theta\tau_{t}^{2}\mathbf{1}).\label{eq:-1-1-1-1}
\end{equation}
Then we evaluate the first input for the threshold function $\eta$:
$\Theta\mathbf{H}^{t\top}\mathbf{r}^{t}+\mathbf{x}^{t}=$
\begin{eqnarray*}
 &  & \Theta\mathbf{H}^{t\top}(\mathbf{H}^{t}\bar{\mathbf{x}}+\mathbf{w}-\mathbf{H}^{t}\mathbf{x}^{t})+\mathbf{x}^{t}\\
 & = & \bar{\mathbf{x}}+\underset{\triangleq\mathbf{u}}{\underbrace{(\Theta\mathbf{H}^{t\top}\mathbf{H}^{t}-\mathbf{I})(\bar{\mathbf{x}}-\mathbf{x}^{t})}}+\underset{\triangleq\mathbf{v}}{\underbrace{\Theta\mathbf{H}^{t\top}\mathbf{w}}.}
\end{eqnarray*}
Regarding term $\mathbf{v}$, it satisfies $\mathbb{V}(v_{i})=\frac{\sigma_{i}^{2}}{m}\times\frac{1}{\sigma_{i}^{4}}\times m\times\sigma^{2}$,
which means $v_{i}\sim\mathrm{N}(0,\sigma^{2}/\sigma_{i}^{2})$. As
for the statistics of term $\bm{u}$, we need the following basic
algebra to measure term $\Theta\mathbf{H}^{t\top}\mathbf{H}^{t}-\mathbf{I}$:

\textit{Suppose that we have two independent Gaussian columns $\mathbf{h}_{i}$
and $\mathbf{h}_{j}$ whose entries are generated from $\mathrm{N}(c,\sigma_{i}^{2}/m)$
and $\mathrm{N}(c,\sigma_{j}^{2}/m)$ respectively. Then $\forall i\neq j$,
we have $\mathbb{E}(\mathbf{h}_{i}^{\top}\mathbf{h}_{j})=mc^{2}$
and $\mathbb{V}(\mathbf{h}_{i}^{\top}\mathbf{h}_{j})=\sigma_{i}^{2}\sigma_{j}^{2}/m+c^{2}(\sigma_{i}^{2}+\sigma_{j}^{2})$.
For $i=j$, we have $\mathbb{E}(\left\Vert \mathbf{h}_{i}\right\Vert ^{2})=mc^{2}+\sigma_{i}^{2}$
and $\mathbb{V}(\left\Vert \mathbf{h}_{i}\right\Vert ^{2})=2\sigma_{i}^{4}/m^{2}+4c^{2}\sigma_{i}^{2}/m$. }

Further denote the covariance matrix of $\bar{\mathbf{x}}-\mathbf{x}^{t}$
as $\mathrm{diag}(\hat{\tau}_{t,1}^{2},...\thinspace,\hat{\tau}_{t,n}^{2})$,
where $\hat{\tau}_{t,i}^{2}=\mathbb{E}|\eta(X+\tau_{t,i}Z,\tau_{t,i}^{2})-X|^{2}$,
$X\sim p_{X}$, $Z\sim\mathrm{N}(0,1)$. Then $\left\{ u_{i}\right\} $
are i.i.d. with zeros mean and variance 
\[
\frac{\hat{\tau}_{t,i}^{2}}{m}+\frac{1}{m\sigma_{i}^{2}}\sum_{j\in[n]}\sigma_{j}^{2}\hat{\tau}_{t,j}^{2},
\]
in which $\frac{\hat{\tau}_{t,i}^{2}}{m}\ll\frac{1}{m\sigma_{i}^{2}}\sum_{j\in[n]}\sigma_{j}^{2}\hat{\tau}_{t,j}^{2}$
and thus negligible. The entry of $\Theta\mathbf{H}^{t\top}\mathbf{r}^{t}+\mathbf{x}^{t}$
can be written as $\bar{x}_{i}+\tau_{t,i}^{t}Z$, where the variance
of $\tau_{t,i}Z=u_{i}+v_{i}$ satisfies 
\begin{align*}
\tau_{t,i}^{2} & =\frac{1}{m\sigma_{i}^{2}}\sum_{j\in[n]}\sigma_{j}^{2}\hat{\tau}_{t,j}^{2}+\frac{\sigma^{2}}{\sigma_{i}^{2}}\\
 & \stackrel{(a)}{=}\frac{1}{m\sigma_{i}^{2}}\sum_{j\in[n]}\sigma_{j}^{2}\mathbb{E}|\eta(X+\tau_{(t-1),j}Z,\tau_{(t-1),j}^{2})-X|^{2}+\frac{\sigma^{2}}{\sigma_{i}^{2}},
\end{align*}
where (a) comes from evaluating the covariance of $\bar{\mathbf{x}}-\mathbf{x}^{t}$. 
\end{IEEEproof}

\section{\label{sec:Proof-of-terProp}Proof of Proposition \ref{prop:powerNepsi}}
\begin{IEEEproof}
Substitute the threshold functions in Lemma \ref{lem:sparseFilter}
to Eq. (\ref{eq:general thres}), it yields 

\begin{equation}
\Psi(\widetilde{\tau}^{2})=\frac{1}{m}\sum_{j\in[n]}\sigma_{j}^{2}\mathbb{E}\left((1-\varepsilon)g_{1}(Z,\widetilde{\tau}^{2})+\varepsilon g_{2}(Z,\widetilde{\tau}^{2})\right)+\sigma^{2},\label{eq:fixedpointEQ}
\end{equation}
 where 
\[
g_{1}(Z,\widetilde{\tau}^{2})=\frac{(1-\varepsilon)/\varepsilon e^{\sigma_{j}^{2}/(2\widetilde{\tau}^{2})}\cosh(Z\sigma_{j}/\widetilde{\tau})+1}{\left((1-\varepsilon)/\varepsilon e^{\sigma_{j}^{2}/(2\widetilde{\tau}^{2})}+\cosh(Z\sigma_{j}/\widetilde{\tau})\right)^{2}},
\]
{\footnotesize{}
\[
g_{2}(Z,\widetilde{\tau}^{2})=\frac{(1-\varepsilon)/\varepsilon e^{\sigma_{j}^{2}/(2\widetilde{\tau}^{2})}\cosh(Z\sigma_{j}/\widetilde{\tau}+\sigma_{j}^{2}/\widetilde{\tau}^{2})+1}{\left((1-\varepsilon)/\varepsilon e^{\sigma_{j}^{2}/(2\widetilde{\tau}^{2})}+\cosh(Z\sigma_{j}/\widetilde{\tau}+\sigma_{j}^{2}/\widetilde{\tau}^{2})\right)^{2}}.
\]
}Since we have

\begin{align*}
\lim_{\widetilde{\tau}^{2}\rightarrow\infty}\Psi(\widetilde{\tau}^{2})= & \frac{1}{m}\sum_{j\in[n]}\sigma_{j}^{2}\left(\frac{1-\varepsilon}{\left(1-\varepsilon\right)/\varepsilon+1}+\frac{\varepsilon}{\left(1-\varepsilon\right)/\varepsilon+1}\right)+\sigma^{2}\\
=\frac{\varepsilon}{m} & \sum_{j\in[n]}\sigma_{j}^{2}+\sigma^{2},
\end{align*}
 one can always tune $\sigma^{2}$ such that $\Psi(\widetilde{\tau}^{2})$
intersects with $f(\widetilde{\tau}^{2})=\widetilde{\tau}^{2}$ and
the point of intersection becomes stable. This point is the highest
one as $\partial\Psi(\widetilde{\tau}^{2})/\partial\widetilde{\tau}^{2}=0$
for all $\widetilde{\tau}^{2}>\varepsilon/m\sum_{j\in[n]}\sigma_{j}^{2}+\sigma^{2}$,
which means $\Psi(\widetilde{\tau}^{2})<\widetilde{\tau}^{2}$ in
this region.
\end{IEEEproof}
\bibliographystyle{IEEEtranMine}
\phantomsection\addcontentsline{toc}{section}{\refname}\bibliography{lib}

\end{document}